\documentclass[aps,prd,notitlepage,twocolumn,showpacs,floatfix]{revtex4} 
\usepackage{graphics}
\usepackage{epsfig}
\usepackage{amsmath}
\usepackage{amssymb}
\usepackage{calc}
\newcommand{\beq}{\begin{equation}}
\newcommand{\eeq}{\end{equation}}
\newcommand{\bea}{\begin{eqnarray}}
\newcommand{\eea}{\end{eqnarray}}
\DeclareMathOperator{\ee}{e}
\DeclareMathOperator{\sgn}{sgn}
\DeclareMathOperator{\Tr}{Tr}

\begin{document}
\title{STU attractors from vanishing concurrence}
\author{P\'eter L\'evay}
\author{Szil\'ard Szalay}
\affiliation{Department of Theoretical Physics, Institute of Physics,
Budapest University of Technology and Economics, H-1521 Budapest, Hungary}
\date{\today}
\begin{abstract}

Concurrence is an entanglement measure characterizing the {\it mixed}
state bipartite correlations inside of a pure state of an
$n$-qubit system. We show that after organizing the charges and
the moduli in the STU model of $N=2$, $d=4$ supergravity to a
three-qubit state, for static extremal spherically symmetric BPS
black hole solutions the vanishing condition for all of the
bipartite concurrences on the horizon is equivalent to the
attractor equations. As a result of this the macroscopic black
hole entropy given by the three-tangle can be reinterpreted as a
linear entropy characterizing the {\it pure} state entanglement 
for an arbitrary bipartite split. Both for
the BPS and non-BPS cases explicit expressions for the
concurrences are obtained, with their vanishing on the horizon is
demonstrated.
\end{abstract}
\maketitle{}

\section{Introduction}
\label{intr}

For STU black holes \cite{Duff1,Behrndt,Belucci} of $N=2$, $d=4$ supergravity the macroscopic black hole entropy
is given by the triality and $\mathrm{SL}(2,\mathbb{Z})^{\otimes 3}$ invariant formula \cite{Behrndt}
\begin{equation}
S_{BH}=\frac{\pi}{G_N}\sqrt{\vert I_4\vert},
\end{equation}
where
\begin{equation}
\begin{split}
I_4(\Gamma)=&
4\sum_{1\leq a<b\leq3}(P^aQ_a)(P^bQ_b)
-(\sum_{I=0}^{3}P^IQ_I)^2\\
&+4Q_0P^1P^2P^3-4P^0Q_1Q_2Q_3.
\end{split}
\label{i4}
\end{equation}
Here $P^I$ and $Q_I$ with $I=0,1,2,3$ are the magnetic and electric charges
characterizing the extremal static spherically symmetric black hole solution
which can be both BPS and non-BPS.
It is also known \cite{Linde,Soroush,Levay,Gimon} that for BPS solutions we have $I_4>0$ and for non-BPS ones
we have $I_4<0$.
As an alternative expression clearly displaying the aforementioned symmetries we have \cite{Duff2}
\begin{equation}
I_4=-4D(\vert\Gamma\rangle)
\end{equation}
\noindent
with $D(\vert\Gamma\rangle)$ is Cayley's hyperdeterminant \cite{Cayley}
of the {\it unnormalized} three-qubit charge state $\vert\Gamma\rangle$
defined as
\begin{equation}
\vert\Gamma\rangle=\sum_{l,k,j=0,1}{\Gamma}_{lkj}\vert lkj\rangle,\qquad
\vert lkj\rangle\equiv\vert l\rangle_3\otimes\vert k\rangle_2\otimes\vert j\rangle_1 \label{chargegamma}
\end{equation}
\noindent
where
\begin{equation}
\begin{split}
\frac{1}{\sqrt{2}}&\begin{pmatrix}
P^0,&P^1,&P^2,&P^3\\
-Q_0,&Q_1,&Q_2,&Q_3
\end{pmatrix}\\
&=\begin{pmatrix}
\Gamma_{000},&\Gamma_{001},&\Gamma_{010},&\Gamma_{100}\\
\Gamma_{111},&\Gamma_{110},&\Gamma_{101},&\Gamma_{011}
\end{pmatrix}.
\end{split}
\label{cgamma}
\end{equation}

On the other hand from quantum information theory it is also known that for an arbitrary {\it normalized}
three-qubit state
\begin{equation}
\vert\psi\rangle=\sum_{l,k,j=0,1}{\psi}_{lkj}\vert lkj\rangle
\end{equation}
the quantity called the {\it three-tangle}
\begin{equation}
{\tau}_{123}\equiv 4\vert D(\vert\psi\rangle)\vert\leq 1
\end{equation}
is a genuine tripartite entanglement measure invariant under $\mathrm{SL}(2, \mathbb{C})^{\otimes 3}$,
a group related to the group of stochastic local operations and classical communication (SLOCC) \cite{CKW,Dur},
and the permutations of the qubits.
Under the full SLOCC group which is $\mathrm{GL}(2, \mathbb{C})^{\otimes 3}$ Cayley's hyperdeterminant $D(\vert\psi\rangle)$ transforms as \cite{Zelevinsky}
\begin{equation}
\label{trafo}
D(\vert\psi\rangle)\mapsto (\det G_3)^2(\det G_2)^2(\det G_1)^2D(\vert\psi\rangle),
\end{equation}
where $G_3\otimes G_2\otimes G_1\in \mathrm{GL}(2, \mathbb{C})^{\otimes 3}$.
Hence for our conventions of Eq.~(\ref{chargegamma}) the STU black hole entropy is
\begin{equation}
S_{BH}=\frac{\pi}{G_N}\sqrt{\tau_{123}(\vert\Gamma\rangle)}.
\label{tangleentropy}
\end{equation}
(Note, that the usual formula appearing in the literature \cite{Duff2,Linde,Levay,Levay3, Borsten} is
$S_{BF}=\frac{\pi}{2G_N}\sqrt{\tau_{123}}$,
however in these studies no $\frac{1}{\sqrt{2}}$ is used in the definition of the charge state of Eq.~(\ref{chargegamma}).)

This interesting correspondence between tripartite entangled systems and stringy black hole solutions has given rise to further results within \cite{stu3,stu4} and outside \cite{e7} the STU context.
This "Black Hole Analogy" (BHA) have repeatedly turned out to be useful for establishing striking results within one of the fields by using methods and results of the other \cite{finite,freud}.
Note however, that the underlying physics (if any) responsible for this black hole-qubit correspondence is still unknown.
At this stage the basic reason for the correspondence seems to be merely that in these two seemingly unrelated fields similar symmetry structures are present.
For example in the very special case of the STU model 
the U-duality group is $\mathrm{SL}(2,\mathbb Z)^{\otimes 3}$ 
and in the three-qubit entanglement case the physically interesting subgroup of SLOCC transformations is $\mathrm{SL}(2,\mathbb C)^{\otimes 3}$.
This observation is the basic reason why entanglement based reformulations of the different aspects of the STU model proved to be useful for providing a quantum information theoretic insight into the entangled web of dualities of the model.
In this respect it is generally believed that the $\mathrm{SL}(2,\mathbb Z)^{\otimes 3}$ and triality invariant expression of Eq.~(\ref{tangleentropy})
should be some sort of macroscopic manifestation of the leading order term of the entanglement entropy for the STU model.

Adopting this view however, we immediately face a problem since the three-tangle making its presence in Eq.~(\ref{tangleentropy}) is {\it not} behaving like {\it entanglement entropy}.
Its physical content in the literature on quantum information
is rather expressed in connection with another property called {\it entanglement monogamy}.
The expression entanglement monogamy is indicating a fact that we cannot share entanglement as a resource for free between the different subsystems.
An equation expressing clearly the physical status of the three-tangle ${\tau}_{123}$ as a residual tangle
is the Coffmann-Kundu-Wootters relation \cite{CKW}
\begin{equation}
{\tau}_{1(23)}(\vert\psi\rangle)=\tau_{123}(\vert\psi\rangle)+
\tau_{12}(\vert\psi\rangle)+\tau_{13}(\vert\psi\rangle)
\label{CKW}
\end{equation}
and similar ones obtained by a permutation of the qubits.
Here
\begin{equation}
\tau_{1(23)}=4\det{\varrho}_1,\qquad {\varrho}_1\equiv\Tr_{23}\vert\psi\rangle\langle\psi\vert
\label{concurrence1}
\end{equation}
\noindent
and $\tau_{13}$ and ${\tau}_{12}$ are the mixed state two-qubit Wootters concurrences squared \cite{Hill,CKW}
associated to the reduced density matrices ${\varrho}_{13}\equiv\Tr_2\vert\psi\rangle\langle\psi\vert$ and
${\varrho}_{12}\equiv\Tr_3\vert\psi\rangle\langle\psi\vert$.
The quantities ${\tau}_{12}$, ${\tau}_{23}$ and ${\tau}_{13}$
are describing the {\it bipartite entanglement} existing {\it within} the tripartite pure state $\vert\psi\rangle$.
On the other hand the quantity $\tau_{1(23)}$ and its cyclically permuted cousins
are known to be directly related to entanglement entropy of the original system characterized by $\vert\psi\rangle$.
This quantity is a bipartite entanglement measure
corresponding to the split $1(23)$.
For the first qubit we have
\begin{equation}
{\tau}_{1(23)}=4\det \varrho_1=2[(\Tr \varrho_1)^2-\Tr\varrho_1^2],
\label{purity}
\end{equation}
\noindent
which is two times the {\it linear entropy} \cite{Beng,Petz} when the state is normalized i.e.~$\Tr\varrho_1=1$.
The linear entropy as defined above is just the so called Tsallis entropy \cite{Beng,Petz} $S_2^{{\rm Tsallis}}$ which is defined for an arbitrary density matrix ${\varrho}$ and $\alpha\in {\mathbb R}^+$ as
\begin{equation}
S_{\alpha}^{{\rm Tsallis}}=\frac{1}{1-{\alpha}}({\rm Tr}\varrho^{\alpha}-1).
\end{equation}\noindent
The quantity ${\rm Tr}{\varrho}^2$ is also occurring in the R\'enyi entropy $S_2$
for an arbitrary $\alpha\in{\mathbb R}^+$ defined as
\begin{equation}
S_{\alpha}^{{\rm Renyi}}=\frac{1}{1-\alpha}\log_2\Tr\varrho^{\alpha}.
\label{renyi}
\end{equation}
\noindent
For normalized states the linear entropy is known to be an approximation
to the von-Neumann entropy which is arising as the $\alpha\to 1$ limit of both the Tsallis and R\'enyi entropies \cite{Beng,Petz}
\begin{equation}
S=-\Tr(\varrho\log_2\varrho).
\end{equation}

These observations indicate that it is the quantity ${\tau}_{1(23)}$
which should be releated to entanglement entropy characterizing directly the {\it bipartite entanglement} corresponding to the split $1(23)$
of {\it some} three-qubit state $\vert\psi\rangle$.
Similar role should be played by the quantities ${\tau}_{2(13)}$ and ${\tau}_{3(12)}$ for the bipartite splits $2(13)$ and $3(12)$.
Though in the black hole context our states are {\it unnormalized} nevertheless based on these considerations
as a simplest candidate for a quantity related to entanglement entropy in the STU context
we can still propose an average of the quantities $\tau_{1(23)}$, $\tau_{2(31)}$ and $\tau_{3(12)}$.
Moreover, due to permutation symmetry of the parties we might even expect a set of equations
${\tau}_{123}(\vert\psi\rangle)={\tau}_{1(23)}(\vert\psi\rangle)={\tau}_{2(13)}(\vert\psi\rangle)={\tau}_{3(12)}(\vert\psi\rangle)$
to hold for some unnormalized "tripartite state" $\vert\psi\rangle$
characterizing the macroscopic configuration.

The simplest choice $\vert\psi\rangle\equiv\vert\Gamma\rangle$ for the underlying state already gives the entropy formula of Eq.~(\ref{tangleentropy}).
However, this clearly fails to be some sort of macroscopic version of an entanglement entropy
since ${\tau}_{12}(\vert\Gamma\rangle)$ and ${\tau}_{13}(\vert\Gamma\rangle)$
generally nonzero hence according to Eq.~(\ref{CKW}) our attempted interpretation ${\tau}_{123}={\tau}_{1(23)}={\tau}_{2(13)}={\tau}_{3(12)}$ fails.

The purpose of the present paper is to show that by employing a three-qubit state
$\vert\Psi(r)\rangle$ which is depending on the charges and the moduli fields,
the latter ones also exhibiting an explicit radial dependence,
our interperetation turns out to be a natural one.
More precisely we will show that for BPS solutions the vanishing condition for the Wootters concurrences
${\tau}_{12}(r)$, ${\tau}_{23}(r)$ and ${\tau}_{13}(r)$ at the black hole horizon $r=0$
is equivalent to the attractor equations used for expressing the moduli in terms of the charges.
The result of this finding is that our desired equation
\begin{equation}
\begin{split}
{\tau}_{123}(\vert\Psi(0)\rangle)
&={\tau}_{1(23)}(\vert\Psi(0)\rangle)\\
&={\tau}_{2(13)}(\vert\Psi(0)\rangle)\\
&={\tau}_{3(12)}(\vert\Psi(0)\rangle),
\end{split}
\label{fontos}
\end{equation}
indeed holds.
Hence a natural interpretation of Eq.~(\ref{tangleentropy}) as an entanglement entropy in a three qubit picture arises.

For readers aware of our previous paper \cite{Levay} this result should not come as a surprise since for {\it double extremal} BPS solutions $\vert\Psi(0)\rangle $ is a GHZ state
for which the Wootters consurrences known to be exactly zero \cite{CKW,Dur}.
However, for such solutions the moduli are constant even away from the horizon
hence Eq.~(\ref{fontos}) holds for $r$ arbitrary.
The novelty here is the demonstration of this result for more general type of solutions for which the off horizon values for quantities like ${\tau}_{1(23)}(r)$ does not satisfy Eq.~(\ref{fontos}).
Displaying an explicit $r$ dependence we will see
how the attractor mechanism unfolds
via forcing the concurrences to be zero at the horizon.
This analysis should be compared with the alternative one based on a discussion of "attractor states" as discussed in our recent paper \cite{levszal}.

The organization of this paper is as follows. In Section II. we
present the background material on the STU model. Section III.
introduces the so called Wootters concurrence an entanglement
measure characterizing the {\it mixed} state two-qubit
correlations inside of an arbitrary $n$-qubit {\it pure} state. We
show that for BPS solutions the vanishing condition of the
concurrence at the horizon is equivalent to the well-known
attractor equations which are usually used to express the
attractor values of the scalar fields in terms of the conserved
charges. Since the attractor flow is essentially the gradient flow
for the BPS mass we also clarify the relationship between the
attractor equations arising from the extremization of the BPS mass
and the same set of equations arising from the vanishing of the
concurrence. In Section IV. by calculating the explicit forms for
the concurrences we demonstrate that the vanishing of these
quantities also holds for the non-BPS flows. Here we first
reconsider the $\frac{1}{2}$-BPS case, then the most general
non-BPS solution with vanishing central charge \cite{Belucci} is
discussed. This section is closed with a discussion on the most
general non-BPS solution with non-vanishing central charge with a
particular emphasis on the $D0-D6$ system. Section V. is devoted
to some geometrical observations connected to the non-BPS case
with non-vanishing central charge. Here we demonstrate that the
vanishing condition for the concurrences is related to the charge
vector being orthogonal to the vector incorporating the moduli, in
a $2+1$ dimensional Minkowski space. This observation enables an
explicit geometric representation for the flat
directions \cite{Gimon,Belucci} occurring in this case. Our
conclusions and some comments are left for Section VI. For the
convenience of the reader we also included an Appendix on the
structure of the BPS mass, now revisited within an entanglement
based three-qubit framework.

\section{The STU model }
\label{conventions}
In the following we consider ungauged $N=2$
supergravity in $d=4$ coupled to    $n$ vector multiplets. The
$n=3$ case corresponds to the $STU$ model. The bosonic part of the
action (without hypermultiplets) is \cite{Behrndt,Duff1}

\begin{equation}
\label{action11}
\begin{split}
{\cal S}=\frac{1}{16\pi}\int d^4x\sqrt{\vert g\vert}
\biggl\{-\frac{R}{2}+G_{a\overline{b}}{\partial}_{\mu}z^a{\partial}_{\nu}{\overline{z}}^{\overline{b}}g^{\mu\nu}&\\
+\left({\rm Im}{\cal N}_{IJ}{\cal F}^I{\cal F}^J+{\rm Re}{\cal N}_{IJ}{\cal F}^I{^\ast{\cal F}^J}\right)&\biggr\}
\end{split}
\end{equation}
Here ${\cal F}^I$, and ${^\ast{\cal F}^I}$,
$I=0,1,2\dots n$ are two-forms associated to the field strengths
${\cal F}^I_{\mu\nu}$  of $n+1$ $\mathrm{U}(1)$ gauge-fields and their
duals.

The $z^a$ $a=1,2\dots n$ are complex scalar (moduli) fields that
can be regarded as local coordinates on a projective special
K\"ahler manifold ${\cal M}$. This manifold for the STU model is
$[\mathrm{SL}(2, \mathbb R)/\mathrm{U}(1)]^{\times 3}$. In the following we will
denote the three complex scalar fields as
\begin{equation}
z^a\equiv x^a-iy^a,\qquad a=1,2,3,\qquad y^a>0.
\end{equation}
With these definitions the metric on the scalar manifold is 
\begin{equation}
G_{a\overline{b}}=\frac{\delta_{a\overline{b}}}{(2y^a)^2}.
\label{targetmetric}
\end{equation}
The metric above can be derived from the K\"ahler potential
\begin{equation}
K= -\log(8y_1y_2y_3)
\label{Kahler}
\end{equation}
as $G_{a\overline{b}}={\partial}_a{\partial}_{\overline{b}}K$.
For the STU model the explicit form the scalar dependent vector
couplings $\nu\equiv{\rm Re}{\cal N}_{IJ}$ and $\mu\equiv{\rm Im}{\cal N}_{IJ}$ can
be found e.g.~in Ref.~\cite{Levay3}.

For the physical motivation of Eq.~(\ref{action11}) we note that
when type IIA string theory is compactified on a $T^6$ of the form
$T^2\times T^2\times T^2$ one recovers $N=8$ supergravity in $d=4$
with $28$ vectors and $70$ scalars taking values in the symmetric
space $\mathrm{E}_{7(7)}/\mathrm{SU}(8)$. This $N=8$ model with an on shell
U-duality symmetry $\mathrm{E}_{7(7)}$ has a consistent $N=2$ truncation
with $4$ vectors and three complex scalars which is just the STU
model. The $D0-D2-D4-D6$ branes wrapping the various $T^2$ give
rise to four electric and four magnetic charges defined as 
\begin{equation}
P^I=\frac{1}{4\pi}\int_{S^2}{\cal F}^I,\qquad
Q_I=\frac{1}{4\pi}\int_{S^2}{\cal G}_I,\quad I=0,1,2,3
\label{ch}
\end{equation}
where
\begin{equation}
{\cal G}_I=\overline{\cal N}_{IJ}{\cal F}^{+I},\qquad 
{\cal F}^{\pm I}_{\mu\nu}={\cal F}^I_{\mu\nu}
\pm \frac{i}{2}{\varepsilon}_{\mu\nu\rho\sigma}{\cal F}^{I\rho\sigma}.
\label{G}
\end{equation}
These charges can be organized into symplectic pairs
\begin{equation}
{\Gamma}\equiv(P^I, Q_J)
\label{Gamma}
\end{equation}
and have units of length. They are related to the
dimensionless quantized charges by some dressing factors.
Normalizing the asymptotic moduli as $y^a(\infty)=1$ and
$x^a(\infty)=B^a$ the dressing factors are essentially the masses
of the underlying branes \cite{Gimon}.

In this paper we are only discussing extremal static spherically
symmetric black hole solutions of the Euler-Lagrange equations of
our Lagrangian of Eq.~(\ref{action11}). For such solutions the
ansatz for the line element is
\begin{equation}
ds^2=-e^{2U(r)}dt^2+e^{-2U(r)}\bigl(dr^2+r^2(d\theta^2+{\sin}^2\theta d{\varphi})\bigr),
\end{equation}
with the warp factor $U(r)$ depending
merely on $r$ which is the distance from the black hole horizon.
After introducing the new variable $\tau\equiv\frac{1}{r}$ now the
dynamics is described by the Lagrangian of a fiducial particle in
a "black-hole potential" $V_{BH}$
\begin{equation}
{\cal L}=\left(\frac{dU}{d\tau}\right)^2
+G_{a\overline{a}}\frac{dz^a}{d\tau}\frac{d\overline{z}^{\overline{a}}}{d\tau}
+e^{2U}V_{BH}(z,\overline{z},P,Q),
\label{eff}
\end{equation}
with the constraint
\begin{equation}
\left( \frac{dU}{d\tau}\right)^2
+G_{a\overline{a}}\frac{dz^a}{d\tau}\frac{d\overline{z}^{\overline{a}}}{d\tau}
-e^{2U}V_{BH}(z,\overline{z},P,Q)=0.
\label{constr}
\end{equation}
Here the black hole potential $V_{BH}$ is depending on the moduli as well on the charges.
Its explicit form is given by
\begin{equation}
V_{BH}=\frac{1}{2}\begin{pmatrix}P^I&Q_I\end{pmatrix}
\begin{pmatrix}(\mu+\nu{\mu}^{-1}\nu)_{IJ}&-(\nu{\mu}^{-1})^J_I\\
-({\mu}^{-1}\nu)^I_J&({\mu}^{-1})^{IJ}\end{pmatrix}
\begin{pmatrix}P^J\\Q_J\end{pmatrix}.
\end{equation}

Extremization of the effective Lagrangian Eq.~(\ref{eff}) with
respect to the warp factor and the scalar fields yields the
Euler-Lagrange equations
\begin{equation}
\ddot{U}=e^{2U}V_{BH},\qquad
\ddot{z}^{a}+\Gamma^a_{bc}\dot{z}^b\dot{z}^c=e^{2U}{\partial}^aV_{BH}.
\end{equation}
In these equations the dots denote derivatives with
respect to $\tau$. These radial evolution equations taken together
with the constraint Eq.~(\ref{constr}) determine the structure of
static spherically symmetric extremal black hole solutions in the
STU model.

As discussed in the introduction it is useful to reorganize the
charges of the STU model into the $8$ amplitudes of a three-qubit
state $\vert\Gamma\rangle$ of Eq.~(\ref{chargegamma}). Notice that
in Eq.~(\ref{chargegamma}) we have introduced the convention of
labelling the qubits from the right to the left. Moreover, for
convenience we have also included a factor $\frac{1}{\sqrt{2}}$
into our definition. The state $\vert\Gamma\rangle$ is a
three-qubit state of a very special kind. First of all this state
defined by the charges need not have to be normalized. Moreover,
the amplitudes of this state are not complex numbers but {\it
real} ones. As a next step we can define a new entangled
three-qubit state $\vert\Psi\rangle$ depending on the charges
$\Gamma$ and also on the moduli  \cite{Levay,Levay3}. This new
state will be a three-qubit state with $8$ complex amplitudes.
However, as we will see it is really a {\it real three-qubit
state}, since it is $\mathrm{U}(2)^{\otimes 3}$ equivalent to a
one with $8$ real amplitudes \cite{Levay,Levay3}.

Now we define the state $\vert\Psi(r)\rangle$ as
\begin{equation}
\begin{split}
&\vert\Psi(z^a,\overline{z}^{\overline{a}},\Gamma)\rangle =\\
&e^{K/2}\begin{pmatrix}\overline{z}^3&-1\\-z^3&1\end{pmatrix}
\otimes \begin{pmatrix}\overline{z}^2&-1\\-z^2&1\end{pmatrix}
\otimes \begin{pmatrix}\overline{z}^1&-1\\-z^1&1\end{pmatrix}\vert\Gamma\rangle.
\end{split}
\label{Psi}
\end{equation}
Here the $r$ dependence is due to the moduli fields
i.e.~$z^a(r)=x^a(r)-iy^a(r)$. Introducing the matrices
\begin{equation}
\begin{split}
{\cal S}_a&\equiv\frac{1}{\sqrt{2y^a}}\begin{pmatrix}\overline{z}^a&-1\\-z^a&1\end{pmatrix}\\
={\cal U}S_a&\equiv\frac{1}{\sqrt{2}}\begin{pmatrix}i&-1\\i&1\end{pmatrix}\frac{1}{\sqrt{y^a}}\begin{pmatrix}y^a&0\\-x^a&1\end{pmatrix},
\end{split}
\label{Smatrix}
\end{equation}
$a=1,2,3$, we have
\begin{equation}
\begin{split}
\vert\Psi(r)\rangle=&({\cal S}_3(r)\otimes {\cal S}_2(r)\otimes {\cal S}_1(r)) \vert\Gamma\rangle\\
=&({\cal U}\otimes {\cal U}\otimes {\cal U})(S_3(r)\otimes S_2(r)\otimes S_1(r))\vert\Gamma\rangle.
\end{split}
\label{3qubitallapot}
\end{equation}
This means that the states $\vert\Psi\rangle$ up to a phase for all values of the moduli
are in the $\mathrm{SL}(2, \mathbb{C})^{\otimes 3}$ orbit of the charge state $\vert\Gamma\rangle$.
Obviously the state $\vert\Psi\rangle$ is an unnormalized three-qubit one
with $8$ complex amplitudes. However, it is {\it not} a genuine complex three-qubit state
but rather a one which is $\mathrm{U}(2)^{\otimes 3}$ equivalent to a real one. This should not come as a surprise
since the symmetry group associated with the STU model is not $\mathrm{SL}(2, \mathbb{C})^{\otimes 3}$
but rather $\mathrm{SL}(2, \mathbb{R})^{\otimes 3}$.
Using these definitions we can write the black hole potential \cite{Belucci,Soroush,Gimon} in the following nice form \cite{Levay3}
\begin{equation}
V_{BH}={\vert\vert\Psi\vert\vert}^2.
\label{Norma}
\end{equation}
Here the norm is defined using the usual scalar product in $\mathbb{C}^8\simeq \mathbb{C}^2\otimes \mathbb{C}^2\otimes\mathbb{C}^2$
with complex conjugation in the first factor.
Since the norm is invariant under $\mathrm{U}(2)^{\otimes 3}$
our choice of the first unitary matrix of Eq.~(\ref{Smatrix}) is not relevant in the structure of $V_{BH}$.
We could have defined a new moduli dependent real state instead of the complex one $\vert\Psi\rangle$
by using merely the $\mathrm{SL}(2, \mathbb{R})$ matrices of Eq.~(\ref{Smatrix}) for their definition.
However, we prefer the complex form of Eq.~(\ref{3qubitallapot}) since it will be useful later.

For computational convenience we use the discrete Fourier (Hadamard) transformed version of our state which is implemented
by acting on $\vert\Psi\rangle$ by $H\otimes H\otimes H$ where
\begin{equation}
H=\frac{1}{\sqrt{2}}\begin{pmatrix}1&1\\1&-1\end{pmatrix}.
\label{Hadamard}
\end{equation}
Hence the Fourier transformed basis states are defined as
\begin{subequations}
\begin{align}
\vert\tilde{0}\rangle&\equiv\frac{1}{\sqrt{2}}(\vert 0\rangle +\vert 1\rangle)=H\vert 0\rangle,\\
\vert\tilde{1}\rangle&\equiv\frac{1}{\sqrt{2}}(\vert 0\rangle -\vert 1\rangle)=H\vert 1\rangle.
\end{align}
\end{subequations}
As a result we get
\begin{equation}
\begin{split}
\vert\tilde{\Psi}(r)\rangle=&(H\otimes H\otimes H)\vert \Psi(r)\rangle\\
=&( {\cal P}\otimes{\cal P}\otimes{\cal P})(S_3(r)\otimes S_2(r)\otimes S_1(r))\vert\Gamma\rangle,
\end{split}
\label{unifourier}
\end{equation}
where
\begin{equation}
{\cal P}=\begin{pmatrix}i&0\\0&-1\end{pmatrix}
\label{phasegate}
\end{equation}
is just $i$ times the usual phase gate from quantum information theory.

\section{The Wootters concurrence and BPS attractors}
\label{WoottersBPS}

For an {\it unnormalized} two-qubit density operator ${\varrho}$
regarded as a nonnegative $4\times 4$ Hermitian matrix acting on the composite Hilbert space
${\cal H}_{AB}={\cal H}_A\otimes{\cal H}_B=\mathbb{C}^2\otimes \mathbb{C}^2$
the Wootters concurrence squared ${\cal C}_{AB}^2$ is defined as \cite{Hill}
\begin{equation}
{\cal C}_{AB}^2\equiv {\tau}_{AB}=[{\rm max}\{0, \lambda_1-\lambda_2-\lambda_3-\lambda_4\}]^2.
\label{Wootters}
\end{equation}
Here $\lambda_1\geq\lambda_2\geq\lambda_3\geq\lambda_4$ are the square-roots of the nonnegative eigenvalues of the matrix
\begin{equation}
{\varrho}\tilde{\varrho}\equiv{\varrho}({\varepsilon}\otimes{\varepsilon}){\varrho}^T({\varepsilon}\otimes{\varepsilon}),
\label{spinflip}
\end{equation}
where ${\varepsilon}$ is the usual $2\times 2$ $\mathrm{SL}(2)$ invariant antisymmetric tensor with
${\varepsilon}_{01}=1$.
We note that for {\it normalized} (i.e.~$\Tr{\varrho}=1$) states we have the extra constraint
$0\leq{\tau}_{AB}\leq 1$ used in quantum information theory.

In the following we will be concerned with calculating the Wootters concurrences ${\tau}_{12}(\vert\Psi(r)\rangle)$,
${\tau}_{23}(\vert\Psi(r)\rangle)$ and ${\tau}_{13}(\vert\Psi(r)\rangle)$
for the {\it unnormalized three-qubit state} of Eq.~(\ref{3qubitallapot}).
Here the subscripts refer to the three different subsystems labelled by the
three different kinds of complex moduli $z^{a}(r)$.
Our aim is to show that the vanishing of these quantities on the horizon
is equivalent to the attractor equations \cite{attractor} for BPS solutions.
A consequence of this is that according to Eq.~(\ref{CKW})
equations (\ref{fontos}) will hold giving rise to the possibility of interpreting the STU black hole entropy as an entanglement entropy.

In order to show this we note that as a byproduct of the Schmidt decomposition
${\varrho}_{23}=\Tr_1\vert\Psi\rangle\langle\Psi\vert$ has merely two nonvanishing eigenvalues
hence $\varrho\tilde{\varrho}$ has two nonvanishing eigenvalues too.
Hence ${\tau}_{23}=(\lambda_1-\lambda_2)^2=\Tr(\varrho_{23}{\tilde{\varrho}}_{23})-2\lambda_1\lambda_2$.
An explicit calculation of these eigenvalues shows that \cite{CKW}
\begin{equation}
{\tau}_{23}=\Tr({\varrho}_{23}\tilde{\varrho}_{23})-\frac{1}{2}\tau_{123}.
\label{ez}
\end{equation}
Now we introduce the notation
\begin{align}
\label{elso}
\Psi_0\equiv\begin{pmatrix}{\Psi}_{000}&{\Psi}_{010}\\{\Psi}_{100}&{\Psi}_{110} \end{pmatrix},\qquad
\Psi_1&\equiv\begin{pmatrix}{\Psi}_{001}&{\Psi}_{011}\\{\Psi}_{101}&{\Psi}_{111} \end{pmatrix},\\
\label{dot}
(\Psi_j\cdot\Psi_k)\equiv\Tr(\Psi_j{\tilde{\Psi}}_k),\quad
{\tilde{\Psi}}_k&\equiv -{\varepsilon}{\Psi}_j^T{\varepsilon}.
\end{align}
Here $j,k=0,1$.
(Recall our convention of labelling the qubits from the right to the left
hence in the case of ${\Psi}_0$ for example the {\it rightmost} i.e.~the {\it first} qubit is $0$.)
In this notation the Wootters concurrence takes the following form
\begin{equation}
\begin{split}
{\tau}_{23}=
  \vert(\Psi_0\cdot\Psi_0)\vert^2
+2\vert(\Psi_0\cdot\Psi_1)\vert^2
+ \vert(\Psi_1\cdot\Psi_1)\vert^2\\
-2\vert [\Psi_0\wedge\Psi_1]^2\vert,
\end{split}
\label{levversion}
\end{equation}
where
\begin{equation}
[\Psi_0\wedge\Psi_1]^2\equiv (\Psi_0\cdot\Psi_0)(\Psi_1\cdot\Psi_1)-(\Psi_0\cdot\Psi_1)^2.
\label{az}
\end{equation}
Notice that $[\Psi_0\wedge\Psi_1]^2$ is just {\it minus} Cayley's hyperdeterminant $-D(\vert\Psi\rangle)$.
Due to the $\mathrm{GL}(2)^{\otimes 3}$ transformation property of this quantity
familiar from Eq.~(\ref{trafo})
and the special structure of Eq.~(\ref{3qubitallapot})
(i.e.~up to phase factors it is on the $\mathrm{SL}(2)^{\otimes 3}$ orbit of the charge state $\vert\Gamma\rangle$)
\begin{equation}
[\Psi_0\wedge\Psi_1]^2=-[\Gamma_0\wedge\Gamma_1]^2,
\end{equation}
i.e.~this quantity is not depending on the moduli.
Hence the moduli dependence of ${\tau}_{23}$ is coming from the first three terms of Eq.~(\ref{levversion}).
However, it is easy to see that these terms are depending merely on the first moduli i.e.~$z^1$.
Indeed the dot product $(A\cdot B)=\Tr(A\tilde B)$ of Eq.~(\ref{dot}) is
an $\mathrm{SL}(2)\times \mathrm{SL}(2)$ invariant one.
For $S_3\otimes S_2\in \mathrm{SL}(2)\times \mathrm{SL}(2)$ the $2\times 2$ matrices $A$ and $B$ transform as
$A\mapsto S_3AS_2^T$ and $B\mapsto S_3BS_2^T$ and the invariance property
$S\varepsilon S^T=\varepsilon$ gives the invariance property of the dot product.
Since up to phase factors $\vert\Psi\rangle$ is the $S_3\otimes S_2\otimes S_1\in \mathrm{SL}(2)^{\otimes 3}$
transformed of the charge state $\vert\Gamma\rangle$
according to these observations the only nontrivial transformation is coming from the factor
of the form $I\otimes I\otimes S_1$ containing merely the moduli $z^1$.
In the following for computational simplicity we will use the Fourier transformed version of our three qubit state
i.e.~Eq.~(\ref{unifourier}). For the first three terms of Eq.~(\ref{levversion}) i.e.~$\Tr(\varrho_{23}{\tilde{\varrho}}_{23})$ we have
\begin{equation}
\begin{split}
\Tr(\varrho_{23}{\tilde{\varrho}}_{23})=&\vert(a\Gamma_0+b\Gamma_1)^2\vert^2+\vert(c\Gamma_0+d\Gamma_1)^2\vert^2\\
&+2\vert(a\Gamma_0+b\Gamma_1)\cdot(c\Gamma_0+d\Gamma_1)\vert^2,
\end{split}
\end{equation}
where
\begin{equation}
{\cal P}S_1=\begin{pmatrix}a&b\\c&d\end{pmatrix}
=\begin{pmatrix}i\sqrt{y}&0\\\frac{x}{\sqrt{y}}&-\frac{1}{\sqrt{y}}\end{pmatrix}.
\end{equation}
Here for simplicity we have used the notation $z\equiv z^1$ and ${\Gamma}_j^2=(\Gamma_j\cdot\Gamma_j)$ etc.
With these definitions we have
\begin{equation}
\begin{split}
&\Tr(\varrho_{23}{\tilde{\varrho}}_{23})=\\
&y^2\vert\Gamma_0^2\vert^2+\frac{1}{y^2}\vert(x\Gamma_0-\Gamma_1)^2\vert^2+2\vert((x\Gamma_0-\Gamma_1)\cdot \Gamma_0)\vert^2.
\end{split}
\end{equation}
After some algebraic manipulations we get
\begin{equation}
\begin{split}
\Tr(\varrho_{23}&{\tilde{\varrho}}_{23})=
-2[\Gamma_0\wedge\Gamma_1]^2\\
+&\frac{1}{y^2}\left[(x^2+y^2)\Gamma_0^2-2x(\Gamma_0\cdot\Gamma_1)+\Gamma_1^2\right]^2.
\end{split}
\end{equation}
Using Eqs.~(\ref{levversion})-(\ref{az}) we obtain
\begin{equation}
\begin{split}
{\tau}_{23}(\vert\Psi\rangle)=
\frac{1}{y^2}\left[(z\Gamma_0-\Gamma_1)\cdot(\overline{z}\Gamma_0-\Gamma_1)\right]^2\\
-2[\Gamma_0\wedge\Gamma_1]^2-2\vert [\Gamma_0\wedge\Gamma_1]^2\vert.
\end{split}
\label{majdnem}
\end{equation}
For BPS black hole solutions $[\Gamma_0\wedge\Gamma_1]^2=-D(\vert\Gamma\rangle)>0$ hence we can write
\begin{equation}
\begin{split}
{\tau}_{23}=
&\left[\frac{1}{y}(z\Gamma_0-\Gamma_1)\cdot(\overline{z}\Gamma_0-\Gamma_1)-2\vert\Gamma_0\wedge\Gamma_1\vert\right]\times\\
&\left[\frac{1}{y}(z\Gamma_0-\Gamma_1)\cdot(\overline{z}\Gamma_0-\Gamma_1)+2\vert\Gamma_0\wedge\Gamma_1\vert\right].
\end{split}
\label{szep}
\end{equation}
Here we have introduced the notation
\begin{equation}
\vert\Gamma_0\wedge\Gamma_1\vert\equiv\sqrt{\Gamma_0^2\Gamma_1^2-(\Gamma_0\cdot\Gamma_1)^2},
\end{equation}
where for BPS solutions the quantity under the square root is positive.
Let us now recall that for $2\times 2$ matrices $A$ and $B$ we have
\begin{align}
\label{osszefuggesek}
\det (A+B)&=\det A+\det B+\Tr(\tilde{A}B),\\
\Tr(A\tilde{A})&=2\det A.
\end{align}
Let us now define the $2\times 2$ matrices
\begin{equation}
\begin{split}
\Lambda_{\pm}(r)\equiv\sqrt{\vert\Gamma_0\wedge\Gamma_1\vert}
\Biggl(
\frac{1}{y(r)}
&\begin{pmatrix}1&x(r)\\x(r)&\vert z(r)\vert^2\end{pmatrix}\\
\pm\frac{1}{\vert\Gamma_0\wedge\Gamma_1\vert}
&\begin{pmatrix}(\Gamma_0\cdot\Gamma_0)&(\Gamma_0\cdot\Gamma_1)\\(\Gamma_0\cdot\Gamma_1)&(\Gamma_1\cdot\Gamma_1)\end{pmatrix}
\Biggr).
\end{split}
\label{matrixok}
\end{equation}
Then using Eq.~(\ref{osszefuggesek}) the Wootters concurrence can be written in the nice form
\begin{equation}
{\tau}_{23}(r)=-\det \Lambda_+(r)\det \Lambda_-(r).
\label{vegso}
\end{equation}
By permutation symmetry of the STU model with the similar looking definitions
\begin{equation}
\begin{split}
\Lambda^{a}_{\pm}(r)\equiv\sqrt{\vert\Gamma_0\wedge\Gamma_1\vert}
\Biggl(
\frac{1}{y^a(r)}
&\begin{pmatrix}1&x^a(r)\\x^a(r)&\vert z^a(r)\vert^2\end{pmatrix}\\
\pm\frac{1}{\vert\Gamma_0\wedge\Gamma_1\vert}
&\begin{pmatrix}(\Gamma_0\cdot\Gamma_0)_a&(\Gamma_0\cdot\Gamma_1)_a\\ (\Gamma_0\cdot\Gamma_1)_a&(\Gamma_1\cdot\Gamma_1)_a\end{pmatrix}
\Biggr),
\end{split}
\label{matrixok2}
\end{equation}
we have
\begin{equation}
{\tau}_{bc}(r)=-\det \Lambda^a_+(r)\det \Lambda^a_-(r),
\label{vegso2}
\end{equation}
where $a,b,c=1,2,3$ with $a,b,c$ different.
Here the dot products like $(\Gamma_0\cdot\Gamma_1)_a$, $a=1,2,3$ refer to the special role the $a$th qubit plays
in building up the relevant $2\times 2$ matrices $\Gamma_0$ and $\Gamma_1$.
Hence for example in the dot product
$(\Gamma_0\cdot\Gamma_1)_2=\Tr(\Gamma_0{\tilde{\Gamma}}_1)$
we have to use the $2\times 2$ matrices
\begin{equation}
\Gamma_0=\begin{pmatrix}\Gamma_{000}&\Gamma_{001}\\\Gamma_{100}&\Gamma_{101}\end{pmatrix},\qquad
\Gamma_1=\begin{pmatrix}\Gamma_{010}&\Gamma_{011}\\\Gamma_{110}&\Gamma_{111}\end{pmatrix}.
\end{equation}

Now we would like to make some observations.
We can write Eq.~(\ref{matrixok}) in the form
\begin{equation}
\Lambda^a_{\pm}=\sqrt{\vert\Gamma_0\wedge\Gamma_1\vert}({\cal M}^a\pm \Gamma^a),
\label{jel2}
\end{equation}
where the $2\times 2$ matrices
\begin{align}
\label{Mmink}
\mathcal{M}^a&=\frac{1}{y^a}
\begin{pmatrix}
1&x^a\\
x^a&(x^a)^2+(y^a)^2
\end{pmatrix},\\
\label{Gmink}
{\Gamma}^a&=
\frac{1}{\vert\Gamma_0\wedge\Gamma_1\vert}
\begin{pmatrix}
(\Gamma_0\cdot\Gamma_0)_a&(\Gamma_0\cdot\Gamma_1)_a\\
(\Gamma_0\cdot\Gamma_1)_a&(\Gamma_1\cdot\Gamma_1)_a
\end{pmatrix}
\end{align}
are having the properties
\begin{equation}
{\cal M}^T={\cal M},\qquad \Gamma^T=\Gamma,\qquad {\cal M},\Gamma\in \mathrm{SL}(2, \mathbb{R}).
\label{tulajdonsagok}
\end{equation}
(For simplicity in the following we supress the $a=1,2,3$ label.)

Now an element $\xi$ of the space of $2\times 2$ {\it real symmetric} matrices can be parametrized as
\begin{equation}
\xi=\begin{pmatrix}T-X&Y\\Y&T+X\end{pmatrix}. \label{minkjel}
\end{equation}
Then this space equipped with the quadratic form
\begin{equation}
\begin{split}
Q:\xi\mapsto Q(\xi)=&-\det (\xi)\\
=&-2(\xi\cdot \xi)\\
=&X^2+Y^2-T^2,
\end{split}
\end{equation}
becomes isomorphic to $2\oplus 1$ dimensional Minkowski space.
The symmetric bilinear form associated to $Q$ is
\begin{equation}
\begin{split}
g:(\xi_1,\xi_2)\mapsto g(\xi_1,\xi_2)=&-\frac{1}{2}\Tr(\xi_1{\tilde{\xi}}_2)\\
=&-\frac{1}{2}(\xi_1\cdot \xi_2)\\
=&X_1X_2+Y_1Y_2-T_1T_2.
\end{split}
\end{equation}
Now in this notation the constraints of Eq.~(\ref{tulajdonsagok}) mean
that ${\cal M}$ and ${\Gamma}$ regarded as vectors in the $2\oplus 1$ dimensional Minkowski space are timelike vectors
lying on the double-sheeted hyperboloid.

In the light of this an alternative form for the expression of the Wootters concurrences squared ${\tau}$ is
\begin{equation}
\begin{split}
&\tau_{bc}(\vert\Psi(r)\rangle)=\\
&\quad\tau_{123}(\vert\Gamma\rangle)[g({\cal M}^a(r),\Gamma^a)+1][g({\cal M}^a(r),\Gamma^a)-1]
\end{split}
\label{minkowskijeloles}
\end{equation}
for $a,b,c$ different.
Note that in this formula the $r$ dependence appears only in ${\cal M}$ containing the moduli.

Hence the general structure of any of our concurrences squared is given by the simple formula $\tau=\tau_{123}(g({\cal M},\Gamma)+1)(g({\cal M},\Gamma)-1)$.
Now it is well-known \cite{Naber} that two timelike vectors are either having the same time orientation with $g({\cal M},\Gamma)<0$, or the opposite one
with $g({\cal M},\Gamma)>0$.
Since $\Tr({\cal M})>0$ due to $y>0$ meaning that the Minkowski vector associated to ${\cal M}$ is {\it future directed} we have the alternatives
\begin{equation}
g({\cal M},\Gamma)<0,\quad \text{i.e.}\quad \Tr({\cal M})>0,\quad \Tr(\Gamma)>0,
\label{alt1}
\end{equation}
or
\begin{equation}
g({\cal M},\Gamma)>0,\quad \text{i.e.}\quad \Tr({\cal M})>0,\quad \Tr(\Gamma)<0.
\label{alt2}
\end{equation}
In either case one of the terms of Eq.~(\ref{minkowskijeloles}) can be made to vanish. So when studying the vanishing conditions for $\tau$,
without the loss of generality we may assume that Eqs.~(\ref{tulajdonsagok}) and (\ref{alt1}) hold,
i.e.~${\cal M}$ and ${\Gamma}$ correspond to future directed timelike vectors lying on the upper sheet of the double-sheeted hyperboloid.
Since we are studying BPS solutions $\Gamma_0^2\Gamma_1^2-(\Gamma_0\cdot\Gamma_1)^2>0$ this means that we may chose charge configurations for which
$\Gamma_0^2>0$ and ${\Gamma}_1^2>0$.

\begin{figure}[!ht]
\setlength{\unitlength}{0.000138\columnwidth}
\begin{picture}(7220,5120)
 \put(0,0){\includegraphics[width=\columnwidth]{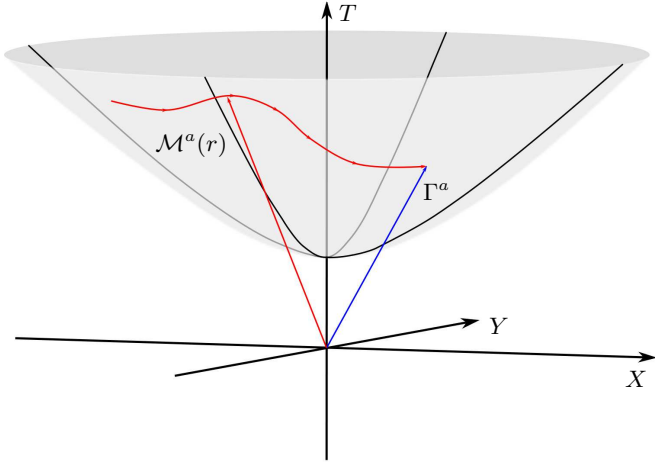}}
  \put(2500,3550){\makebox(0,0)[r]{\strut{}$\mathcal{M}^a(r)$}}
  \put(5000,3000){\makebox(0,0)[r]{\strut{}$\Gamma^a$}}
  \put(7200, 900){\makebox(0,0)[r]{\strut{}$X$}}
  \put(5650,1500){\makebox(0,0)[r]{\strut{}$Y$}}
  \put(3950,5000){\makebox(0,0)[r]{\strut{}$T$}}
 \end{picture}
 \caption{Illustration of the BPS-attractor flow.
The moduli (represented by $\mathcal{M}^a(r)$ of
Eq.~(\ref{Mmink})) converge to their horizon-values (represented
by the $\Gamma^a$s of Eq.~(\ref{Gmink})). The corresponding
Minkowski vectors are defined by using Eq.~(\ref{minkjel}).}
 \label{fig_hypbps}
\end{figure}

Now the upper sheet of the double sheeted hyperboloid is a model
for the hyperbolic plane. The bilinear form $g$ restricts to a
Riemannian metric on tangent spaces to the hyperboloid. The
distance $d({\cal M},\Gamma)$ with respect to this metric between
two points ${\cal M}$ and $\Gamma$ on the hyperboloid is given by
the expression.
\begin{equation}
\cosh^2d({\cal M},\Gamma)=[g({\cal M},\Gamma)]^2.
\label{metric}
\end{equation}
According to Eq.~(\ref{minkowskijeloles}) ${\tau}$ is vanishing precisely when
the distance between ${\cal M}$ and $\Gamma$ is zero, i.e.~${\cal
M}=\Gamma$. However, generally ${\cal M}$ is depending on the
radial coordinate but the matrix $\Gamma$ is constant. On the
horizon we have the simultaneous vanishing condition of all
Wootters concurrences ${\cal M}^{(a)}(0)=\Gamma^{(a)}$ i.e.
\begin{equation}
\begin{split}
\frac{1}{y^a(0)}
&\begin{pmatrix}1&x^a(0)\\x^a(0)&(x^a(0))^2+(y^a(0))^2\end{pmatrix}\\
&=\frac{1}{\vert\Gamma_0\wedge\Gamma_1\vert}
\begin{pmatrix}(\Gamma_0\cdot\Gamma_0)_a&(\Gamma_0\cdot\Gamma_1)_a\\(\Gamma_0\cdot\Gamma_1)_a&(\Gamma_1\cdot\Gamma_1)_a\end{pmatrix}.
\label{attractor}
\end{split}
\end{equation}
(See in Fig.~\ref{fig_hypbps}.)
These are precisely the BPS
attractor equations that can be written in the more familiar
form \cite{Behrndt}
\begin{equation}
z^a(0)=x^a(0)-iy^a(0)
=\frac{(\Gamma_0\cdot\Gamma_1)_a+i\vert\Gamma_0\wedge\Gamma_1\vert}{(\Gamma_0\cdot\Gamma_0)_a}.
\label{usual}
\end{equation}
Here the negativity of the imaginary
part ensures the positivity of the K\"ahler potential. In
conclusion: the Wootters concurrences squared ${\tau}_{ab}(r)$
$a,b=1,2,3$ , $a\neq b$ of our three-qubit state
$\vert\Psi(r)\rangle$ are vanishing on the horizon
(${\tau}_{ab}(0)=0$) {\it precisely} when the BPS attractor
equations hold. Solutions for which Eq.~(\ref{attractor}) holds
for all values of $r$ are the {\it double extremal solutions}.
However, for more general type of solutions Eq.~(\ref{attractor})
does not hold for $r\neq 0$, hence the Wootters concurrences
generally not zero away from the horizon. We will see examples for
this phenomenon in the next section.

The important corollary of our result is that according to the
CKW-relations of Eq.~(\ref{CKW}) the entanglement entropies
${\tau}_{a(bc)}(r)$ having generally a different value off the
horizon, will flow to the same value, namely
${\tau}_{123}(\vert\Gamma\rangle)$.
Hence equations
\begin{equation}
{\tau}_{a(bc)}(0)=\tau_{123}(0)=\tau_{123}(\vert\Gamma\rangle)
\label{conclusion}
\end{equation}
for $a,b,c$ distinct will indeed hold.

Now recall that the BPS attractor flow is essentially the gradient
flow of the BPS mass hence as a next step it is a natural question
to ask what is the relationship between this flow and the flow
obtained from our considerations related to the Wootters
concurrences. The value of the BPS mass squared is obtained by
putting the asymptotic values for the moduli into the $r$
dependent formula \cite{Duff1,Levay}
\begin{equation}
\begin{split}
M_{BPS}^2= \frac{1}{4}\langle\Gamma\vert\Bigl(
{\cal N}_3\otimes &{\cal N}_ 2\otimes{\cal N}_1\\
-{\cal N}_3\otimes\varepsilon\otimes\varepsilon
-\varepsilon \otimes&{\cal N}_2\otimes\varepsilon
-\varepsilon\otimes\varepsilon \otimes{\cal N}_1
\Bigr)\vert\Gamma\rangle.
\end{split}
\end{equation}
Here
\begin{equation}
{\cal N}_a\equiv{\cal M}_a^{-1}
={\tilde{\cal M}}_a
=\frac{1}{y^a}\begin{pmatrix}(x^a)^2+(y^a)^2&-x^a\\-x^a&1\end{pmatrix}
\end{equation}
(for $a=1,2,3$) is an $\mathrm{SL}(2,\mathbb{R})$ matrix, where
the $r$-dependence of the moduli is left implicit and the extra
factor of $\frac{1}{4}$ is partly arising from our unusual
normalization used in Eq.~(\ref{chargegamma}). In the Appendix it
is shown that by attaching a special role to one of the qubits
(e.g.~to the first one) this formula can be written in the
following form
\begin{equation}
M_{BPS}^2=\det ({\cal Z}_+)
\end{equation}
where
\begin{equation}
\begin{split}
{\cal
Z}_{\pm}\equiv\frac{1}{\sqrt{8}}\sqrt{\vert\gamma_0\wedge\gamma_1\vert}
\Biggl( \frac{1}{y}
&\begin{pmatrix}1&x\\x&\vert z\vert^2\end{pmatrix}\\
\pm\frac{1}{\vert\gamma_0\wedge\gamma_1\vert}
&\begin{pmatrix}(\gamma_0\cdot\gamma_0)&(\gamma_0\cdot\gamma_1)\\
(\gamma_0\cdot\gamma_1)&(\gamma_1\cdot\gamma_1)
\end{pmatrix}\Biggr).
\end{split}
\label{zpm}
\end{equation}
Here
\begin{align}
\gamma_{i\mu}&={\Sigma}_{\mu\nu}{\Gamma}_{i\nu},\\
\Sigma&\equiv{\cal N}_3\otimes\varepsilon+\varepsilon\otimes {\cal N}_2,\\
\gamma_i&={\cal N}_3{\Gamma}_i{\varepsilon}^T+{\varepsilon}{ \Gamma}_i{\cal N}_2^T.
\end{align}
In these expressions we can regard
${\gamma}_{i\mu}$, $i=0,1$, $\mu=1,2,3,4$ as a pair of four-vectors
or a pair of $2\times 2$ matrices depending on the charges and the
moduli $z^2$ and $z^3$. Alternatively one can regard
$\gamma_{kji}(z^2,z^3,P^I,Q_I)$ as a three-qubit state
displaying {\it no} dependence on $z^1\equiv z=x-iy$.

Now the attractor equations \cite{Behrndt} fixing the values of the
moduli at $r=0$ are coming from the extremization of $M_{BPS}$.
Employing the shorthand notation $z^a(0)\equiv z^a=x^a-iy^a$ these
equations can be rewritten as
\begin{equation}
\begin{split}
\frac{1}{y}&\begin{pmatrix}1&x\\x&x^2+y^2\end{pmatrix}\\
&=\frac{1}{\vert\Gamma_0\wedge\Gamma_1\vert}
\begin{pmatrix}
(\Gamma_0\cdot\Gamma_0)_1&(\Gamma_0\cdot\Gamma_1)_1\\
(\Gamma_0\cdot\Gamma_1)_1&(\Gamma_1\cdot\Gamma_1)_1
\end{pmatrix}\\
&=\frac{1}{\vert\gamma_0\wedge\gamma_1\vert}
\begin{pmatrix}
(\gamma_0\cdot\gamma_0)_1&(\gamma_0\cdot\gamma_1)_1\\
(\gamma_0\cdot\gamma_1)_1&(\gamma_1\cdot\gamma_1)_1
\end{pmatrix}.
\end{split}
\label{attractor2}
\end{equation}
Clearly due to the triality symmetry of the STU model these are equivalent to Eq.~(\ref{attractor}).

Notice that the $4\times 4$ matrices ${\cal Z}_{\pm}$ are similar
in structure to the ones showing up in recent investigations on
domain walls and issues of marginal stability of $N=4$ dyons
\cite{bpsmass,dyons}. This is of course not a coincidence since
using duality transformations the $N=2$ STU model can be related
to such $N=4$ models in a number of different ways \cite{Duff1}.
One possible way to see the correspondence between such $N=2$ and
$N=4$ structures is to consider a toroidal $T^6$ compactification
of the heterotic string. Indeed let us label by $x^{\mu},
\mu=0,1,2,3$ the noncompact coordinates, and restrict attention to
merely a subsector of the theory in which we include only those
gauge fields that are associated with the $4\mu$ and $5\mu$
components of the ten dimensional metric and antisymmetric tensor
field. Moreover, let us consider merely the scalar fields coming
from the $mn$, $4\leq m,n\leq 5$ components of such fields
i.e.~the ones associated with one of the tori $T^2$. In this
subsector the $T$-duality transformations as elements of
$\mathrm{O}(2,2,{\mathbb Z})$ are acting on the charges and the
moduli \cite{Duff1,bpsmass}. The moduli fields can now be included
into a $4\times 4$ matrix
\begin{equation}
M=\begin{pmatrix}G^{-1}&G^{-1}B\\-BG^{-1}&G-BG^{-1}B\end{pmatrix},\qquad
L=\begin{pmatrix}0&I\\I&0\end{pmatrix},
\end{equation}
where
\begin{equation}
G=\frac{y^3}{y^2}\begin{pmatrix}\vert z^2\vert^2&x^2\\x^2&1\end{pmatrix},\qquad
B=\begin{pmatrix}0&x^3\\-x^3&0\end{pmatrix}.
\end{equation}
These matrices satisfy the constraints
\begin{equation}
M^T=M,\qquad MLM^T=L.
\end{equation}

Now after employing the $SO(4)$ matrix
\begin{equation}
W=\begin{pmatrix}0&0&1&0\\0&0&0&-1\\0&1&0&0\\1&0&0&0\end{pmatrix}
\end{equation}
we obtain the result
\begin{equation}
W(M-L)W^T={\cal N}_3\otimes {\cal
N}_2-\varepsilon\otimes\varepsilon. \label{l}
\end{equation}
Clearly the transformation based on $W$ exploits the group isomorphism
$\mathrm{O}(2,2)\simeq \mathrm{SL}(2,{\mathbb R})\times \mathrm{SL}(2,{\mathbb R})$.
Now using Eq.~(\ref{l}) and the results of the appendix (see
Eq.~(\ref{g}) in this respect) it is easy to demonstrate the
structural similarity of the usual expressions for the BPS mass
\cite{Duff1,bpsmass,dyons} and our expressions used in the context
of the STU model. In this picture our $1+2$ split of the qubits
with the first one playing a distinguished role corresponds to the
split of the $U$-duality group to $S$ and $T$-duality
transformations in the form 
$\mathrm{SL}(2,{\mathbb Z})\times \mathrm{O}(2,2,{\mathbb Z})$.

Notice also that ${\rm det}{\cal Z_+}\equiv \vert Z_+\vert^2$ and
${\rm det}{\cal Z_-}\equiv \vert Z_-\vert^2$ are just the
magnitudes squared of the eigenvalues of the matrix $ZZ^{\dagger}$
formed from the canonical form of the $4\times 4$ antisymmetric
central charge matrix $Z$ of the $N=4$ supersymmetry algebra. As
it is well-known the largest eigenvalue in theories with $N=4$
supersymmetry plays the role of the BPS mass (just like the single
central charge plays the same role in $N=2$ theories).

It is interesting to compare this role played by the matrices
${\cal Z}_{\pm}$ of Eq.~(\ref{zpm}) with the role played by the
corresponding ones ${\Lambda}_{\pm}$ of Eq.~(\ref{matrixok})
governing the structure of the Wootters concurrence. In both cases
the determinants of these matrices define important quantities.
According to Eq.~(\ref{vegso}) the determinants of
${\Lambda}_{\pm}$ are giving rise to an expression for the
concurrence squared. On the other hand the determinant of one of
${\cal Z}_{\pm}$ gives the expression for the BPS mass squared.
Due to the attractor mechanism the moduli are stabilized at the
horizon. During this process the value of the BPS mass flows to a
value related to the macroscopic black hole entropy as
\begin{equation}
S_{BH}=\frac{\pi}{G_N} M_{BPS}^2(0)
=\frac{\pi}{G_N}\vert Z_+(0)\vert^2
=\frac{\pi}{G_N}\sqrt{\tau_{123}(\vert\Gamma\rangle)}.
\end{equation}
On the other hand the concurrence flows to a
vanishing value, giving rise to an interpretation of this black
hole entropy as a linear entanglement entropy in this three-qubit
picture as displayed by Eq.~(\ref{conclusion}).

\section{Explicit expressions for the Wootters concurrences}
\label{Szilard}

In this section we will give explicit expressions for the Wootters
concurrences calculated for all three classes of non-degenerate
attractor flows of the STU-model. These classes are  as follows:
$\frac{1}{2}$-BPS, non-BPS $Z=0$, non-BPS $Z\neq 0$. We consider
the most general configurations \cite{Belucci}, when all the
charges are switched on. We will show, that in all three cases the
Wootters concurrences squared vanish at the event-horizon.

In the following we use the $p^I$, $q_I$ quantized charges instead
of the $P^I$, $Q_I$ dressed ones. This is because we would like to
use the most general non-BPS $Z\neq0$ solution \cite{Belucci} which
has been produced by using an U-duality transformation acting on
such quantized charges. The dressed charges are rescaled
quantities related to the quantized (undressed) ones via factors
coming from the asymptotic volume moduli. During this rescaling,
our important quantities transform with the powers of the $d=4$
Newton constant $G_N$. It turns out \cite{levszal} that $\Psi$
scales with $\sqrt{G_N}$, i.e.~$\vert\Psi(P^I,Q_I,z^a)\rangle =
\sqrt{G_N}\vert\Psi(p^I,q_I,\tilde{z}^a)\rangle$ where
$\tilde{z}^a$ are the moduli calculated with the $p^I$, $q_I$
quantized charges. Because of this, the Wootters concurrences
squared transform with $G_N^2$
\begin{equation}
\tau_{bc}(P^I,Q_I,z^a)= G_N^2\tau_{bc}(p^I,q_I,\tilde{z}^a),
\end{equation}
which can be seen most easily from Eq.~(\ref{levversion}).

In this section we denote  the undressed three-qubit charge state as $\Gamma$,
defined similarly as in Eq.~(\ref{cgamma}).

\subsection{The most general $\frac{1}{2}$-BPS solution}
\label{halfBPS}
Let us start with the supersymmetric case.
The general solution of the attractor flow equations is
\begin{subequations}
\begin{align}
\ee^{-4U(r)} &= I_4(\mathcal{H}(r)),\\
\tilde{z}^a(r)&=
\frac{h^a(r)+i\partial_{h_a}\sqrt{I_4(\mathcal{H}(r))}}
{h^0(r)+i\partial_{h_0}\sqrt{I_4(\mathcal{H}(r))}},
\end{align}
\end{subequations}
where where $U(r)$ is the warp factor and $\mathcal{H}(r)$ can be constructed from the harmonic functions
\begin{subequations}
\begin{align}
h^I(r) &= \overline{p}^I + p^I \frac{1}{r},\\
h_I(r) &= \overline{q}_I+ q_I \frac{1}{r},
\end{align}
\end{subequations}
similarly to $\Gamma$ in Eq.~(\ref{cgamma}). With these quantities
we can write
$\mathcal{H}(r)= \overline{\Gamma} + \Gamma \frac{1}{r}$.
Two constraints have to be satisfied by $\overline{\Gamma}$:
\begin{equation}
I_4(\overline{\Gamma})=1,\qquad
\langle \Gamma , \overline{\Gamma}\rangle = p^I \overline{q}_I -q_J \overline{p}^J = 0.
\end{equation}
The black hole charge configurations supporting the $\frac{1}{2}$-BPS attractors at the event horizon
are the ones satisfying the following set of constraints \cite{Belucci}
\begin{equation}
I_4(\Gamma)>0,\qquad
p^ap^b-p^0q_c>0.
\label{bpsfeltetelek}
\end{equation}
The same holds for the whole flow, i.e.
\begin{equation}
I_4(\mathcal{H})>0,\qquad
h^ah^b-h^0h_c>0.
\label{bpsfeltetelek2}
\end{equation}
In this case the moduli can be rewritten as
\begin{equation}
\label{md2}
\tilde{z}^a=
\frac{h^I h_I-2h^ah_a-i\sqrt{I_4(\mathcal{H})}}
{2(h^bh^c-h^0h_a)},
\end{equation}
where $r$-dependence is implicit, and summation on $I$ is
understood. The indices $a,b,c$ are distinct elements of the set
$\{1,2,3\}$, and no summation  on $a$ is implied. As in Eq.~(\ref{elso})
we can define the $2\times 2$ matrices
$\mathcal{H}_0$ and $\mathcal{H}_1$. We have already seen that in
$\tau_{bc}$ only the $a$th moduly appears. In the following we
indicate by the notation $(\mathcal{H}_0\cdot\mathcal{H}_0)_a$
which qubit for the construction of $\mathcal{H}_0$ and
$\mathcal{H}_1$ plays a special role. Now the solutions for the
moduli take the following form
\begin{equation}
\tilde{z}^a=
\frac{(\mathcal{H}_0\cdot \mathcal{H}_1)_a+i\vert \mathcal{H}_0\wedge\mathcal{H}_1 \vert}
{(\mathcal{H}_0\cdot \mathcal{H}_0)_a }.
\end{equation}
By virtue of (\ref{majdnem}), with $[\Gamma_0\wedge\Gamma_1]^2>0$ we obtain
\begin{equation}
\begin{split}
&\tau_{bc}(\Psi)=-4[\Gamma_0\wedge\Gamma_1]^2\\
&+\frac{1}{(\tilde{y}^a)^2}
\left[ \vert \tilde{z}^a\vert^2 (\Gamma_0\cdot\Gamma_0)_a -2\tilde{x}^a (\Gamma_0\cdot\Gamma_1)_a + (\Gamma_1\cdot\Gamma_1)_a \right]^2,
\end{split}
\end{equation}
which can be cast to the form
\begin{equation}
\label{tbps}
\begin{split}
&\tau_{bc}(\Psi)
=\frac{1}{ [\mathcal{H}_0\wedge\mathcal{H}_1]^2 }
\bigl[
   (\mathcal{H}_1\cdot\mathcal{H}_1)_a(\Gamma_0\cdot\Gamma_0)_a\\
&-2(\mathcal{H}_0\cdot\mathcal{H}_1)_a(\Gamma_0\cdot\Gamma_1)_a
  +(\mathcal{H}_0\cdot\mathcal{H}_0)_a(\Gamma_1\cdot\Gamma_1)_a
\bigr]^2\\
&-4[\Gamma_0\wedge\Gamma_1]^2.
\end{split}
\end{equation}
As we have proven in Section \ref{WoottersBPS} $\tau_{bc}$ must vanish at the
horizon. One can also see this directly from the expression above.
Indeed, since $\lim_{r\to0}r\mathcal{H}_0=\Gamma_0$ and
$\lim_{r\to0}r\mathcal{H}_1=\Gamma_1$ it follows, that the first
term of Eq.~(\ref{tbps}) converges to the second one and then
$\lim_{r\to0}\tau_{bc}=0$.

Consider now the special $D0-D4$ case.
For this solution only the charges of $D0$ and $D4$-branes are turned on
which can be obtained by putting $p^0=0$ and $q_a=0$ into our general expression.
This results in
$(\Gamma_0\cdot\Gamma_0)_a = -p^bp^c$,
$(\Gamma_0\cdot\Gamma_1)_a = 0$,
$(\Gamma_1\cdot\Gamma_1)_a = -q_0p^a$,
and
\begin{equation}
\begin{split}
&\tau_{bc}(\Psi)
=-4q_0p^ap^bp^c\\
+&\frac{1}{ [\mathcal{H}_0\wedge\mathcal{H}_1]^2 }
\left[
 (\mathcal{H}_1\cdot\mathcal{H}_1)_ap^bp^c
 -(\mathcal{H}_0\cdot\mathcal{H}_0)_aq_0p^a
\right]^2.
\end{split}
\end{equation}
We can further specialise this by noticing that this case admits
axion-free attractor-flows, i.e.~moduli with vanishing real part.
(The asymptotic limit of the real part of the moduli gives the
B-fields realized on the tori, so these flows does not admit
non-trivial B-fields.) To implement this case $h^0$ and $h_a$ must
be zero as it can be seen from Eq.~(\ref{md2}), so we have to switch
off also $\overline{p}^0$ and $\overline{q}_a$. The Wootters
concurrences squared for this axion-free case are
\begin{equation}
\tau_{bc}(\Psi)
=\frac{1}{h_0h^ah^bh^c}
\left[ h_0h^ap^bp^c -q_0p^ah^bh^c \right]^2.
\end{equation}

\subsection{The most general non-BPS $Z=0$ solution}
\label{nBPSZ0}
The non-BPS $Z=0$ solutions \cite{Belucci} can be obtained from  the $\frac{1}{2}$-BPS ones
by simply changing the sign of any two imaginary parts of the moduli.
This leaves the (\ref{Kahler}) K\"ahler potential invariant
and yields the following change for the (\ref{bpsfeltetelek}) $\frac{1}{2}$-BPS constraints
\begin{equation}
\begin{split}
I_4>0,\qquad
&p^ap^b-p^0q_c>0,\\
&p^bp^c-p^0q_a<0,\\
&p^cp^a-p^0q_b<0.
\end{split}
\label{nbpsZnullfeltetelek}
\end{equation}
During the calculation of $\vert \tilde{\Psi}\rangle$ the moduli only appear in the $S_a$ matrices.
We can carry out the sign flip of some $\tilde{y}^a$ by employing the Pauli matrix $\sigma_3$
\begin{equation}
-\sigma_3 S_a=
\frac{1}{\sqrt{\tilde{y}^a}}\begin{pmatrix}
-\tilde{y}^a&0\\
-\tilde{x}^a&1
\end{pmatrix}.
\end{equation}
Due to this observation in order to calculate ${\tau}_{bc}$ we
have to use $(\sigma_3\otimes\sigma_3\otimes
I)\vert\tilde{\Psi}\rangle$ etc or
$(\sigma_1\otimes\sigma_1\otimes I) \vert\Psi\rangle$ etc. Since
these states are local-unitary equivalent to the original ones,
the $\tau_{bc}$'s are of the same form as in the $\frac{1}{2}$-BPS
case.

\subsection{The most general non-BPS $Z\neq0$ solution}
\label{nBPSZn0}
The calculation of the Wootters concurrences squared
is not straightforward in this case,
due to the complicated expressions for the attractor flow \cite{Belucci},
so we show the main steps.

The non-BPS $Z\neq0$ attractor flow is
\begin{subequations}
\begin{align}
\ee^{-4U(r)} &= h_0(r)h_1(r)h_2(r)h_3(r)-b^2,\\
\label{nBPSaf}
\tilde{x}^a(r) &=\frac{\varsigma_a\nu_a^2C^a_1+(\varsigma_a-\varrho_a)\nu_aC^a_2-\varrho_aC^a_3}
{\nu_a^2C^a_1+2\nu_aC^a_2 +C^a_3},\\
\tilde{y}^a(r) &=\frac{(\varsigma_a+\varrho_a)2\nu_aC_4 }
{\nu_a^2C^a_1+2\nu_aC^a_2 +C^a_3},
\end{align}
\end{subequations}
where
\begin{subequations}
\begin{align}
\nu_a &=\nu \ee^{\alpha_a},\\
\nu&=\left(\frac{2p^1p^2p^3 - p^0p^I q_I + \sqrt{-[\Gamma_0\wedge\Gamma_1]^2} p^0}
{2p^1p^2p^3 - p^0p^I q_I - \sqrt{-[\Gamma_0\wedge\Gamma_1]^2} p^0}\right)^{\frac{1}{3}},\\
\label{vsig}
\varsigma_a &=\frac{ (\Gamma_0\cdot\Gamma_1)_a-\sqrt{-[\Gamma_0\wedge\Gamma_1]^2 } }{ (\Gamma_0\cdot\Gamma_0)_a },\\
\label{vrho}
\varrho_a &=\frac{-(\Gamma_0\cdot\Gamma_1)_a-\sqrt{-[\Gamma_0\wedge\Gamma_1]^2 } }{ (\Gamma_0\cdot\Gamma_0)_a },
\end{align}
\end{subequations}
are charge-dependent constants.
The $\alpha_a$ real constants satisfying the constraint
$\alpha_1+\alpha_2+\alpha_3=0$,
account for the flat directions \cite{Gimon,Belucci}.
The harmonic functions are now defined as
\begin{equation}
h_I (r)= b_I+(-I_4(\Gamma))^{\frac{1}{4}}\frac{1}{r},
\end{equation}
giving rise to the $r$-dependent quantities
\begin{subequations}
\begin{align}
\label{nX1}
C^a_1&=h_bh_c+h_0h_a+2b,\\
C^a_2&=h_bh_c-h_0h_a,\\
C^a_3&=h_bh_c+h_0h_a-2b,\\
\label{nX4}
C_4&=\ee^{-2U} = \sqrt{h_0h_1h_2h_3-b^2},
\end{align}
\end{subequations}
also making their presence in Eqs.~(\ref{nBPSaf}).
Using the (\ref{vsig}) and (\ref{vrho}) form of $\varsigma_a$ and $\varrho_a$
we can separate the terms containing $h_I(r)$ in the moduli, as
\begin{subequations}
\begin{align}
\label{xX}
\tilde{x}^a(r)&= \frac{(\Gamma_0\cdot\Gamma_1)_a}{(\Gamma_0\cdot\Gamma_0)_a}
-\frac{\sqrt{-[\Gamma_0\wedge\Gamma_1]^2 }}{(\Gamma_0\cdot\Gamma_0)_a}C^a_x(r),\\
\label{yY}
\tilde{y}^a(r)&=
-\frac{\sqrt{ -[\Gamma_0\wedge\Gamma_1]^2}}{(\Gamma_0\cdot\Gamma_0)_a}C^a_y(r).
\end{align}
\end{subequations}
Here the only $r$ dependent terms are
\begin{subequations}
\begin{align}
C^a_x(r)&=\frac{\nu_a^2C^a_1-C^a_3}{\nu_a^2C^a_1+2\nu_aC^a_2 +C^a_3},\\
C^a_y(r)&=\frac{4\nu_aC_4}{\nu_a^2C^a_1+2\nu_aC^a_2 +C^a_3}.
\end{align}
\end{subequations}

After this preparation recall (\ref{majdnem}) for this case, when $[\Gamma_0\wedge\Gamma_1]^2<0$
\begin{equation}
\begin{split}
&\tau_{bc}(\Psi)=\\
&\frac{1}{(\tilde{y}^a)^2}\left[ \vert \tilde{z}^a\vert^2 (\Gamma_0\cdot\Gamma_0)_a -2\tilde{x}^a (\Gamma_0\cdot\Gamma_1)_a + (\Gamma_1\cdot\Gamma_1)_a \right]^2.
\end{split}
\end{equation}
Straightforward calulation then shows
\begin{equation}
\tau_{bc}(\Psi)
=-[\Gamma_0\wedge\Gamma_1]^2  \frac{1}{(C^a_y)^2}\left[ (C^a_x)^2+(C^a_y)^2-1 \right]^2.
\end{equation}
If we notice that $C^a_1C^a_3-(C^a_2)^2=4(C_4)^2$ (see Eqs.~(\ref{nX1})-(\ref{nX4}))
then it turns out that
\begin{equation}
(C^a_x)^2+(C^a_y)^2=\frac{\nu_a^2C^a_1-2\nu_a C^a_2 + C^a_3}{\nu_a^2 C^a_1+2\nu_a C^a_2 + C^a_3},
\end{equation}
and with this
\begin{equation}
\label{jippii}
\begin{split}
\tau_{bc}(\Psi)
=&-[\Gamma_0\wedge\Gamma_1]^2\left[ \frac{C^a_2}{C_4} \right]^2\\
=&\frac{-I_4(\Gamma)}{4}\ee^{4U}\left[h_0h_a-h_bh_c\right]^2.
\end{split}
\end{equation}
We note that this expression is independent of the flat directions of the non-BPS $Z\neq0$ flow.

Now we can show easily that $\tau_{bc}$ vanishes at the event horizon,
by calculate its $r\to 0$ limit.
Indeed since
\begin{equation}
\tau_{bc}(\Psi)
=\frac{-I_4(\Gamma)}{4}\frac{\left[b_0b_a-b_bb_c + (\dots)\frac{1}{r} \right]^2}
{b_0b_1b_2b_3-b^2 + \dots -I_4(\Gamma)\frac{1}{r^4}},
\end{equation}
it follows that
\begin{equation}
\lim_{r\to0}\tau_{bc} = 0.
\end{equation}
By virtue of this calculation one can see that also in this non BPS situation
a GHZ-like state with vanishing Wootters-concurrence have been distilled at the event-horizon.

In the asymptotically Minkowski region the warp factor must be equal to $1$  hence
\begin{equation}
\begin{split}
\lim_{r\to\infty}\tau_{bc}
=&\frac{-I_4(\Gamma)}{4}\left[b_0b_a-b_bb_c\right]^2\\
=&\frac{-I_4(\Gamma)}{4}\left[(b_0b_a)^2+(b_bb_c)^2 -2(1+b^2)\right].
\end{split}
\end{equation}

As a special subcase of this non-BPS $Z\neq0$ one now we consider the $D0-D6$ solution.
This charge-configuration independently of the signs of the charges can appear only in the non-BPS regime
because $I_4 = -(p^0q_0)^2 < 0$.
Originally, the general solution have been constructed using an
$\mathrm{SL}(2,\mathbb{R})^{\otimes 3}$ U-duality transformation of this $D0-D6$ solution \cite{Gimon,Belucci}.
Due to the $\varsigma_a$, $\varrho_a$ parametrisation of this transformation,
(see in Eqs.~(\ref{vsig}) and (\ref{vrho}))
we can not produce neither the identity transformation
nor the transformations that bring us back to the $D0-D6$ case with different charges.
Hence we can not simply write the $D0-D6$ charges into the corresponding formulae for the horizon-limit of $\tau_{bc}$.
However, for the calculation of $\tau_{bc}$ on the horizon
we can proceed by directly using the original $D0-D6$ solutions \cite{Belucci}:
\begin{subequations}
\begin{align}
\ee^{-4U(r)} &= h_0(r)h_1(r)h_2(r)h_3(r)-b^2,\\
\tilde{x}^a(r)&=\frac{\nu_a' C^a_2}{C^a_3},\\
\tilde{y}^a(r)&=\frac{\nu_a'2 C_4}{C^a_3},
\end{align}
\end{subequations}
with the notation introduced in Eqs.~(\ref{nX1})-(\ref{nX4}).
Here
\begin{equation}
\nu'_a =\ee^{\alpha_a}\left(\frac{q_0}{p^0}\right)^{\frac1 3}.
\end{equation}
Recall Eq.~(\ref{majdnem}) for this case, when $[\Gamma_0\wedge\Gamma_1]^2<0$.
For the $D0-D6$ charge configuration
$-2(\Gamma_0\cdot\Gamma_1)_a=p^0q_0$,
$(\Gamma_0\cdot\Gamma_0)_a=(\Gamma_1\cdot\Gamma_1)=0$,
and the expression
\begin{equation}
\begin{split}
\tau_{bc}(\Psi) &=\left[ \frac{\tilde{x}^a}{\tilde{y}^a}p^0q_0 \right]^2\\
&= \frac{-I_4(\Gamma)}{4}\ee^{4U}\left[h_0h_a-h_bh_c\right]^2
\end{split}
\end{equation}
is the same as in Eq.~(\ref{jippii}) of the general non-BPS $Z\neq0$
case. The $D0-D6$ solution also supports an axion-free attractor
flow. This case is obtained if $\tilde{x}^a=0$ for all $a$ or
equivalently when $h_0h_a=h_bh_c$. This holds if and only if
$h_0=h_a=h_b=h_c$ and in this case $\tau_{bc}(\Psi)=0$ for all
values of $r$ along the flow.

\section{ Some geometrical observations on the non-BPS $Z\neq0$ case }
\label{Szilard2}
By means of a geometric approach in Section \ref{WoottersBPS} we have shown
that the vanishing condition for the Wootters concurrence on the event horizon
is equivalent to the $\frac1 2$-BPS attractor equations.
Then in Section \ref{halfBPS} we have illustrated this equivalence by calculating the explicit
$r$-dependent expressions for the concurrence and taking the horizon limit.
In the non-BPS $Z\neq0$ case, however,
we could not relate the vanishing of the concurrence on the event horizon
 to the non-BPS $Z\neq0$ attractor equations.
In Section \ref{nBPSZn0} we have merely demonstrated the vanishing of ${\tau}_{bc}$ by means of an explicit calculation using the known solution.
In this section we investigate the geometrical aspects of the vanishing of the concurrence even for this case.

In the non-BPS $Z\neq0$ case we have $I_4(\Gamma)=4[\Gamma_0\wedge\Gamma_1]^2<0$
and an expression similar to Eq.~(\ref{minkowskijeloles}) for the concurrence
can be obtained
\begin{equation}
\tau_{bc}=\tau_{123}(\Gamma)[g(\mathcal{M}^a(r),\Gamma^a)]^2.
\end{equation}
However, this time the determinant of the $2\times 2$ matrix
\begin{equation}
\Gamma^a=
\frac{1}{\sqrt{-[\Gamma_0\wedge\Gamma_1]^2}}
\begin{pmatrix}
(\Gamma_0\cdot\Gamma_0)_a&(\Gamma_0\cdot\Gamma_1)_a\\
(\Gamma_1\cdot\Gamma_0)_a&(\Gamma_1\cdot\Gamma_1)_a
\end{pmatrix}
\end{equation}
is $-1$.
As a result of this the vector
corresponding to $\Gamma^a$ is {\it spacelike} and lies on the one-sheeted hyperboloid.

\begin{figure}[!ht]
 \setlength{\unitlength}{0.000138\columnwidth}
 \begin{picture}(7220,5120)
  \put(0,0){\includegraphics[width=\columnwidth]{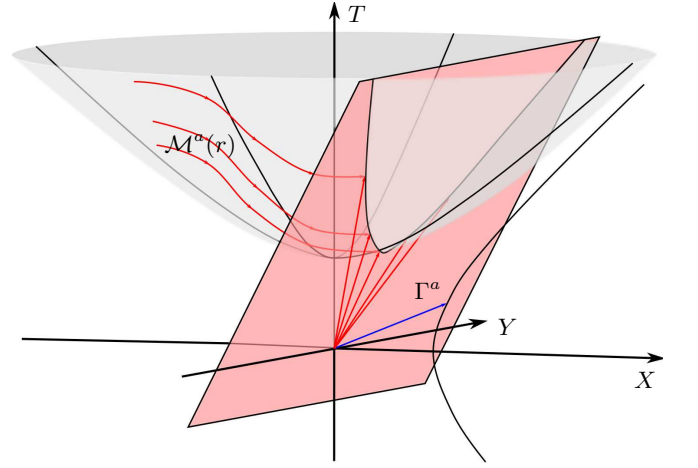}}
  \put(2500,3550){\makebox(0,0)[r]{\strut{}$\mathcal{M}^a(r)$}}
  \put(4800,1900){\makebox(0,0)[r]{\strut{}$\Gamma^a$}}
  \put(7200, 900){\makebox(0,0)[r]{\strut{}$X$}}
  \put(5650,1500){\makebox(0,0)[r]{\strut{}$Y$}}
  \put(3950,5000){\makebox(0,0)[r]{\strut{}$T$}}
 \end{picture}
 \caption{Illustration of the non-BPS $Z\neq0$ flow.
Now the moduli (represented by $\mathcal{M}^a(r)$ of
Eq.~(\ref{Mmink})) converge to their horizon-values which are not
fixed uniquely by the vanishing of the Wootters-concurrence. The
vanishing condition in this case merely forcing $\mathcal{M}^a(0)$
to be Minkowski-orthogonal to $\Gamma^a$.}
 \label{fig_hyplane}
\end{figure}

Now the vanishing condition for $\tau_{bc}$ on the horizon is equivalent to the one
of ortogonality of $\mathcal{M}^a(0)$ and $\Gamma^a$.
However, the condition
\begin{equation}
\lim_{r\to0}g(\mathcal{M}^a(r),\Gamma^a)=0
\end{equation}
does not fix $\mathcal{M}^a(0)$ uniquely. The attractor value of
$\mathcal{M}$ in this non-BPS $Z\neq0$ case is an element of a
one-parameter submanifold of the upper sheet of the double sheeted
hyperboloid. Indeed it is a geodesic obtained via intersecting
this hyperboloid with a plane containing the origin.
(See Fig.~\ref{fig_hyplane}.)

In the following we show that this freedom in the value of $\mathcal{M}^a$ on the horizon
is related to the flat directions appearing for the non-BPS $Z\neq0$ flow.
Using Eqs.~(\ref{Mmink}), (\ref{Gmink}) and (\ref{minkjel}) we can write the components
of the Minkowski vectors corresponding to the $2\times 2$ symmetric matrices  $\Gamma^a$ and $\mathcal{M}^a$ as
\begin{align}
\Gamma^a&\longmapsto \frac{1}{\sqrt{-[\Gamma_0\wedge\Gamma_1]^2}}
\begin{pmatrix}
\frac{1}{2}((\Gamma_1\cdot\Gamma_1)_a-(\Gamma_0\cdot\Gamma_0)_a)\\
(\Gamma_0\cdot\Gamma_1)_a\\
\frac{1}{2}((\Gamma_1\cdot\Gamma_1)_a+(\Gamma_0\cdot\Gamma_0)_a)
\end{pmatrix},\\
\mathcal{M}^a&\longmapsto \frac{1}{\tilde{y}^a}
\begin{pmatrix}
\frac{1}{2}(|\tilde{z}^a|^2-1)\\
\tilde{x}^a\\
\frac{1}{2}(|\tilde{z}^a|^2+1)
\end{pmatrix}.
\end{align}
In the following we refer to these vectors also as $\Gamma^a$ and $\mathcal{M}^a$.

First we consider the special case when the vector corresponding
to $\Gamma^a$ is $(1,0,0)$. Now $(\Gamma_0\cdot\Gamma_1)_a = 0$
and $(\Gamma_0\cdot\Gamma_0)_a =-(\Gamma_1\cdot\Gamma_1)_a$. This
charge configuration is arising for the non-BPS $Z\neq0$ $D0-D4$
or $D2-D6$ system. Then $\mathcal{M}^a(0)$ lies on the hyperbola
parametrized as $(0,\sinh\mu,\cosh\mu)$. Hence we see that the
simplest candidate for the parametrization of such
$\mathcal{M}^a(0)$ is given by the rapidity $\mu$. On the other
hand from the explicit form of the corresponding solution we know
that the second coordinate of $\mathcal{M}^a(0)$ should be
$\frac{\tilde{x}^a}{\tilde{y}^a}$ which now equals to $\frac{C^a_x}{C^a_y}$.
(See Eqs.~(\ref{xX})-(\ref{yY}).)
The horizon limit of this coordinate is
$\sgn(\nu)\sinh(\varphi+\alpha_a)$, and the full vector
corresponding to $\mathcal{M}^a(0)$ is
\begin{equation}
\mathcal{M}^a(0)=\begin{pmatrix}
0\\
\sgn(\nu)\sinh(\varphi+\alpha_a)\\
\cosh(\varphi+\alpha_a)
\end{pmatrix},
\end{equation}
where $\varphi=\ln|\nu|$.
Here the parameter $\alpha_a$ for the flat directions of the flow
makes its presence, and we can identify the other parameter $\mu$
as $\sgn(\nu)(\varphi+\alpha_a)$.

As a next step we can allow more general charges by lifting
$\Gamma^a$ vertically from the horizontal plane.
A Lorenz boost implementing this change leaves the second
coordinate of $\mathcal{M}^a(0)$ and $\Gamma^a$ and also the role
played by the flat directions invariant. Finally a rotation around
the third coordinate axis does not cause any new effect, hence we
can conclude that the freedom in the value of $\mathcal{M}^a$ on
the horizon is caused by the appearance of flat directions in the
attractor flow.

\section{Conclusions}

In this paper we have demonstrated how an entanglement based
picture related to the properties of a charge and moduli-dependent
three-qubit state $\vert\Psi\rangle$ already used in our previous
papers \cite{Levay,levszal} can produce further insight on issues
concerning the attractor mechanism. Our main calculational tool
was a new quantity which until now has only been used in quantum
information theory: the Wootters concurrence. This quantity which
is related to the important concept of the entanglement of
formation  \cite{Hill} can be used to characterize the mixed two
qubit correlations inside of an arbitrary n-qubit pure state. In
the case of the STU model of $N=2$, $d=4$ supergravity we made use
of this quantity for the $n=3$ case. For static spherically
symmetric extremal black hole solutions from $\vert\Psi\rangle$ a
one parameter family of three-qubit states $\vert\Psi(r)\rangle$
is emerging. It is obtained as the 
$\mathrm{SL}(2,{\mathbb R})^{\times 3}$
orbit of a suitable three-qubit charge state
$\vert\Gamma\rangle$. This means that now via the moduli this
state also exhibits an implicit $r$ dependence where $r$ is the
radial coordinate measuring the distance from the horizon.

We have demonstrated that for $\frac{1}{2}$-BPS solutions the
vanishing conditions for the Wootters concurrences at $r=0$ are
{\it equivalent} to the usual attractor equations. For non-BPS
black holes we merely proved the vanishing of the concurrences via
an explicit calculation employing the explicit form of such
solutions \cite{Belucci,Soroush,Gimon}. As it is well-known the
attractor equations make it possible to express the values of the
moduli fields at the horizon in terms of the charges facilitating
a calculation of the macroscopic black hole Bekenstein-Hawking
(BH) entropy. As it was shown \cite{Duff2} for the STU model this
quantity can be reinterpreted as a one related to the unique
triality and SLOCC invariant \cite{Dur} tripartite entanglement
measure, the three-tangle, for the charge state
$\vert\Gamma\rangle$. However, within the theory of quantum
entanglement the physical meaning of the three-tangle of
$\vert\Gamma\rangle$ is {\it not} related to the desirable notion
of {\it entanglement entropy} but rather to a related concept
called {\it entanglement monogamy}. This latter term can be used
as a meme referring to the fact that entanglement cannot be shared
for free. For three-qubit systems this sharing of entanglement is
characterized by the Coffmann-Kundu-Wootters (CKW) relations
\cite{CKW} encapsulating the precise form of the trade-off between
the parties of the tripartite system.

Now in order to retain the meaning of the black hole entropy as
some sort of entanglement entropy even within this three-qubit
entanglement based scenario we have an interesting possibility. To
uncover this possibility let us provide an alternative
characterization of the particular class of STU black hole
solutions as follows. This new characterization will be based not
on the properties of the charge states \cite{Linde}
$\vert\Gamma\rangle$ but on the properties of the "attractor
states" \cite{levszal} $\vert\Psi(0)\rangle$. Notice that unlike
the charge state $\vert\Gamma\rangle$ the attractor state
$\vert\Psi(0)\rangle$ is a quantity of {\it dynamics} as it is
reflecting the end point of the attractor flow in moduli space.
 In this case according to the CKW
inequalities the vanishing conditions for {\it all} of the
concurrences squared of $\vert\Psi(r)\rangle$ for $r=0$ makes it
also possible to reinterpret the macroscopic black hole entropy as
a linear entropy for $\vert\Psi(r)\rangle$ at the horizon arising
from an {\it arbitrary} single partite-bipartite split of our
tripartite system.

In this three-qubit picture the physical meaning of the STU black
hole entropy as a linear entropy is as follows.
When attaching to
{\it one} of our qubits (e.g.~to the first one)
 a distinguished role this split of roles in a particular duality frame corresponds to
a split of a subgroup of the $U$-duality group to $S$ and
$T$-dualities answering the group structure 
$\mathrm{SL}(2,\mathbb Z)\times \mathrm{O}(2,2,\mathbb Z)$.
This group structure now refers to the
admissible set of local operations. In the stringy context a local
operation of that kind can be for example the change of the string
coupling corresponding to the change in the dilaton $\Phi^1=\log
y^1$. As it is well-known the linear entropy is zero precisely
when the three-qubit state $\vert\Psi(0)\rangle$ can be
transformed by such local manipulations to a product state of the
form $\vert\varphi\rangle_{32}\otimes \vert\chi\rangle_1$ or
$\vert\gamma\rangle_3\otimes\vert\beta\rangle_2\otimes\vert\alpha\rangle_1$
where the subscripts refer to the particular subsystems
corresponding to the qubits. This case corresponds to a class of
small black holes \cite{Linde} i.e.~ones that have vanishing
macroscopic BH entropy. (However, they can have nonzero terms
arising from higher curvature corrections.)

Now notice that due to triality symmetry as displayed by
Eq.~(\ref{fontos}) if one of the linear entropies is vanishing for
$\vert\Psi(0)\rangle$ the same holds for {\it all of them}. An
important consequence of this is that the attractor states with
vanishing BH entropy are {\it totally separable} (at least in our
approximation neglecting quantum corrections). This means that
they are on the orbit of the local group $\mathrm{SL}(2,\mathbb Z)^{\otimes3}$
of a state of the form
$\vert\gamma\rangle_3\otimes\vert\beta\rangle_2\otimes\vert\alpha\rangle_1$.
Since according to Eq.~(\ref{fontos}) at $r=0$ all of the linear
entropies are the same in the following we will refer to {\it the}
linear entropy or {\it the} BH entropy of the "state"
$\vert\Psi(0)\rangle$.

 Now a value of the BH entropy different from
zero indicates the impossibility of transforming our attractor
state to the fully separable product form by {\it "local"}
operations. In order to do this we have to employ a set of {\it
"global"} transformations belonging to a larger group containing
the local one as a subgroup, i.e.~possibly the full $U$ duality
group or the group of Peccei-Quinn transformations \cite{Peccei}
incorporating the sub-leading quantum perturbative corrections to
the cubic special geomety of the STU model \cite{PQ}. The latter
group of transformations is capable of transforming small black
holes to large ones \cite{PQ} by including also in the family of
admissible operations Witten theta-shifts \cite{Witten,PQ,corr}.
The particular value of the linear entropy is then related to the
macroscopic black hole entropy of a large \cite{Linde} black hole,
i.e.~to a one having nonzero BH entropy. These black holes contain
some amount of entanglement as measured by the linear entropy of
the state $\vert\Psi(0)\rangle$. In the quantum information
theoretic context in order to transform their states to the
product form manipulations belonging to the full {\it nonlocal}
group of transformations are needed. In the stingy black hole
context the precise form of the corresponding statement referring
to some property of the attractor state $\vert\Psi(0)\rangle$ in
connection with the action of the full $U$-duality or Peccei-Quinn
group should be clarified.

 Clearly at this stage
it is not at all clear whether the entanglement entropy of the
"attractor state" $\vert\Psi(0)\rangle$ can be regarded (and in
what sense) as a macroscopic manifestation of the entanglement
entropy of a genuine quantum state whose degeneracy is responsible
for the black hole entropy. Until this important issue is
clarified the entanglement contained in $\vert\Psi(r)\rangle$
merely provides a nice way of describing the web of dualities of
the STU model using the techniques of quantum information.

\section{Acknowledgement}
This work was supported by the New Hungary Development Plan
(Project ID: T\'AMOP-4.2.1/B-09/1/KMR-2010-0002).

\appendix

\section{BPS mass formula}
\label{appA}
The BPS mass squared is \cite{Duff1}
\begin{equation}
\begin{split}
M_{BPS}^2= \frac{1}{4}\langle\Gamma\vert\Bigl(
{\cal N}_3\otimes &{\cal N}_ 2\otimes{\cal N}_1\\
-{\cal N}_3\otimes\varepsilon\otimes\varepsilon
-\varepsilon \otimes&{\cal N}_2\otimes\varepsilon
-\varepsilon\otimes\varepsilon \otimes{\cal N}_1
\Bigr)\vert\Gamma\rangle.
\end{split}
\label{BPSmass}
\end{equation}
Here
\begin{equation}
{\cal N}_a\equiv{\cal M}_a^{-1}={\tilde{\cal
M}}_a=\frac{1}{y^a}\begin{pmatrix}(x^a)^2+(y^a)^2&-x^a\\-x^a&1\end{pmatrix}
\end{equation}
is an $\mathrm{SL}(2,\mathbb{R})$ matrix,
where as usual $\tilde{\cal M}=-\varepsilon{\cal M}^T\varepsilon$.

Let us define the quantity
\begin{equation}
{\Pi}_{\pm}\equiv I\otimes I\pm{\cal N}_3\varepsilon\otimes{\cal N}_2\varepsilon.
\label{projector}
\end{equation}
Then it is easy to check that the
$4\times 4$ matrices $\frac{1}{2}{\Pi}_{\pm}$ are rank two projectors i.e.
\begin{subequations}
\begin{align}
{\Pi}^2_{\pm}&=2{\Pi}_{\pm},&\qquad
{\Pi}_{\pm}{\Pi}_{\mp}&=0,\\
{\Pi}_+ +{\Pi}_-&=2I\otimes I,&\qquad
{\Pi}_{\pm}^T&={\Pi}_{\pm}.
\label{projtul}
\end{align}
\end{subequations}
We will make special use of ${\Pi}_-$ hence we adopt the notation
\begin{equation}
{\Pi}\equiv{\Pi}_-.
\end{equation}
We will also need the quantity
\begin{equation}
\Sigma\equiv{\cal N}_3\otimes\varepsilon+\varepsilon\otimes{\cal N}_2
\label{Sigma}
\end{equation}
with the properties
\begin{subequations}
\begin{align}
{\Sigma}^T&=-{\Sigma},&\qquad
{\Pi}_-{\Sigma}&={\Pi}{\Sigma}=2{\Sigma},\\
{\Pi}_{+}{\Sigma}&=0,&\qquad
\Sigma\tilde{\Sigma}&=-2{\Pi}
\label{sigtul}
\end{align}
\end{subequations}
where
$\tilde{\Sigma}=\varepsilon\otimes\varepsilon\Sigma^T\varepsilon\otimes\varepsilon$.
With these results we can rewrite the BPS mass squared with
special role attached to the first qubit as
\begin{equation}
M_{BPS}^2=\frac{1}{4}\langle\Gamma\vert {\cal L}\otimes{\cal
N}_1\vert\Gamma\rangle-\frac{1}{4}\langle\Gamma\vert\Sigma\otimes\varepsilon\vert\Gamma\rangle,
\label{newBPSmass}
\end{equation}
where
\begin{equation}
{\cal L}\equiv
-{\Pi}g=-g\tilde{\Pi}={\cal N}_3\otimes{\cal N}_2-\varepsilon\otimes\varepsilon,\quad
g\equiv\varepsilon\otimes\varepsilon.
\label{g}
\end{equation}
In this form of the BPS mass squared only the first term contains the
moduli $z^1$ associated to our first qubit with special status.

With the special role for the first qubit we can represent the $8$ charges
either as a pair of four-vectors,
\begin{equation}
{\Gamma}_{0\mu}=\begin{pmatrix}\Gamma_{000}\\ \Gamma_{010}\\ \Gamma_{100}\\ \Gamma_{110}\end{pmatrix},\qquad
{\Gamma}_{1\mu}=\begin{pmatrix}\Gamma_{001}\\ \Gamma_{011}\\ \Gamma_{101}\\ \Gamma_{111}\end{pmatrix}
\end{equation}
where $\mu=1,2,3,4$, or a pair of $2\times 2$ matrices
\begin{equation}
\Gamma_{0jk}=\begin{pmatrix}\Gamma_{000}&\Gamma_{010}\\\Gamma_{100}&\Gamma_{110}\end{pmatrix},\quad
\Gamma_{1jk}=\begin{pmatrix}\Gamma_{001}&\Gamma_{011}\\\Gamma_{101}&\Gamma_{111}\end{pmatrix}
\end{equation}
where $j,k=0,1$.

Now we adopt the definition
\begin{equation}
\gamma_{i\mu}={\Sigma}_{\mu\nu}{\Gamma}_{i\nu},\qquad
\gamma_i={\cal N}_3{\Gamma}_i{\varepsilon}^T+{\varepsilon}{\Gamma}_i{\cal N}_2^T.
\label{fura}
\end{equation}
In this definition we can regard ${\gamma}_i$, $i=0,1$ as a pair of four-vectors
or a pair of $2\times 2$ matrices
depending on the charges and the moduli $z^2$ and $z^3$.
Alternatively we can regard $\gamma_{kji}(z^2,z^3,P^I,Q_I)$
as a three-qubit state displaying {\it no} dependence on $z^1$.

Now as a first property of $\gamma_i$ one can check that
\begin{equation}
{\gamma}_i({\cal N}_2\varepsilon)^T=({\cal N}_3\varepsilon){\gamma}_i.
\label{felhasznal}
\end{equation}
The second property of $\gamma_i$ we need is the vanishing of the commutator
\begin{equation}
[{\cal N}_3\varepsilon,\gamma_1{\tilde{\gamma}}_0]=0.
\label{commutator}
\end{equation}

Now as a trick to give a new look to Eq.~(\ref{newBPSmass}) we note that
\begin{equation}
2{\cal L}=\Sigma g\Sigma,\qquad
4\Sigma=-\Sigma\tilde{\Sigma}\Sigma.
\label{trick}
\end{equation}
Using this with $z^1\equiv z$ we have
\begin{equation}
\begin{split}
\frac{1}{4}\langle\Gamma\vert&{\cal L}\otimes{\cal N}_1\vert\Gamma\rangle=\\
&\frac{1}{8y}\left(\vert
z\vert^2(\gamma_0\cdot\gamma_0)-2x(\gamma_0\cdot\gamma_1)+(\gamma_1\cdot\gamma_1)\right),
\end{split}
\label{elsotag}
\end{equation}
where as usual
\begin{equation} (A\cdot
B)=A^{\mu}g_{\mu\nu}B^{\nu}=\Tr(A\tilde{B}), \label{cdot2} \end{equation}
with $g$ known from Eq.~(\ref{g}).

For the second term of Eq.~(\ref{newBPSmass}) we have
\begin{equation}
\begin{split}
2\langle\Gamma\vert\Sigma\otimes\varepsilon\vert\Gamma\rangle
&=-\gamma_{0\mu}{\tilde{\Sigma}}_{\mu\nu}\gamma_{1\nu}\\
&=-\Tr({\cal N}_3\varepsilon\gamma_1{\tilde{\gamma}}_0+{\tilde{\gamma}}_0\gamma_1({\cal N}_2\varepsilon)^T)\\
&= -2\Tr({\cal N}_3\varepsilon\gamma_1{\tilde{\gamma}}_0),
\end{split}
\label{masodiktag}
\end{equation}
where in the last equality we have used
Eq.~(\ref{felhasznal}). In order to transform this term further we
use the identity
\begin{equation}
[\Tr(A)]^2-\Tr(A^2)=2\det (A),
\label{id}
\end{equation}
valid for $2\times 2$ matrices. By virtue of this we have
\begin{equation}
[\Tr({\cal N}_3\varepsilon\gamma_1{\tilde{\gamma}}_0)]^2 =\Tr(
{\cal N}_3\varepsilon\gamma_1{\tilde{\gamma}}_0{\cal
N}_3\varepsilon\gamma_1{\tilde{\gamma}}_0)+2\det(\gamma_1{\tilde{\gamma}}_0).
\end{equation}
Using in the first term the commutation property of
Eq.~(\ref{commutator}) and the identity ${\cal
N}_3\varepsilon{\cal N}_3={\cal N}_3\varepsilon{\cal
N}_3^T=\varepsilon$ we get
\begin{equation}
[\Tr({\cal N}_3\varepsilon\gamma_1{\tilde{\gamma}}_0)]^2
=-\Tr(\gamma_1{\tilde{\gamma}_0})^2+2\det(\gamma_1{\tilde{\gamma}}_0).
\end{equation}
Now in the first term
of this we use once again Eq.~(\ref{id}) and recall the definition
of Eq.~(\ref{az}) to obtain
\begin{equation}
\begin{split}
[\Tr({\cal N}_3\varepsilon\gamma_1{\tilde{\gamma}}_0)]^2
&=-[\Tr(\gamma_1{\tilde{\gamma}}_0)]^2+[2\det(\gamma_0)][2\det (\gamma_1)]\\
&= [\gamma_0\wedge\gamma_1]^2.
\end{split}
\label{Cayleyuj}
\end{equation}
These considerations yield the formula for the BPS mass squared
with the first qubit playing a special role
\begin{equation}
M_{BPS}^2=\frac{1}{8}\left[\frac{1}{y}(z\gamma_0-\gamma_1)
\cdot(\overline{z}\gamma_0-\gamma_1)+2\vert\gamma_0\wedge\gamma_1\vert\right],
\label{legujabb}
\end{equation}
where
\begin{equation}
\vert\gamma_0\wedge\gamma_1\vert=\sqrt{(\gamma_0\cdot\gamma_0)(\gamma_1\cdot\gamma_1)-(\gamma_0\cdot\gamma_1)^2},
\end{equation}
and recall also Eqs.~(\ref{Sigma}), (\ref{fura}) and
(\ref{cdot2}). Now after using the definitions of Eq.~(\ref{zpm})
from this we get the final result
\begin{equation}
M_{BPS}^2=\det ({\cal Z}_+). \label{centralcharge}
\end{equation}
These considerations are to be compared with the ones
connected to Eqs.~(\ref{matrixok}) and (\ref{vegso}) and the ones
of Section \ref{WoottersBPS}.

\end{document}